\def\getvaxfonts{\message{Installing Computer Modern Fonts.}
\font\fourteenrm=cmr12 scaled 1200
\font\twelverm=cmr12
\font\tenrm=cmr10
\font\ninerm=cmr9
\font\eightrm=cmr8
\font\sevenrm=cmr7
\font\sixrm=cmr6
\font\fiverm=cmr5
\font\fourteeni=cmmi12 scaled 1200
\font\twelvei=cmmi12
\font\teni=cmmi10
\font\ninei=cmmi9
\font\eighti=cmmi8
\font\seveni=cmmi7
\font\sixi=cmmi6
\font\fivei=cmmi5
\font\fourteensy=cmsy10 scaled 1440
\font\twelvesy=cmsy10 scaled 1200
\font\tensy=cmsy10
\font\ninesy=cmsy9
\font\eightsy=cmsy8
\font\sevensy=cmsy7
\font\sixsy=cmsy6
\font\fivesy=cmsy5
\font\fourteenex=cmex10 scaled 1440
\font\twelveex=cmex10 scaled 1200
\font\tenex=cmex10
\font\fourteenbf=cmbx12 scaled 1200
\font\twelvebf=cmbx12
\font\tenbf=cmbx10
\font\ninebf=cmbx9
\font\eightbf=cmbx8
\font\sevenby=cmbx7
\font\sixbf=cmbx6
\font\fivebf=cmbx5
\font\fourteenit=cmti12 scaled 1200
\font\twelveit=cmti12
\font\tenit=cmti10
\font\nineit=cmti9
\font\eightit=cmti8
\font\sevenit=cmti7
\font\fourteensl=cmsl12 scaled 1200
\font\twelvesl=cmsl12
\font\tensl=cmsl10
\font\ninesl=cmsl9
\font\eightsl=cmsl8
\font\fourteentt=cmtt12 scaled 1200
\font\twelvett=cmtt12
\font\tentt=cmtt10
\font\ninett=cmtt9
\font\eighttt=cmtt8
\font\fourteensc=cmcsc10 scaled 1440
\font\twelvesc=cmcsc10 scaled 1200
\font\tensc=cmcsc10
\font\fourteenss=cmss12 scaled 1200
\font\twelvess=cmss12
\font\tenss=cmss10
\font\niness=cmss9
\font\eightss=cmss8
\font\fourteenssl=cmssi12 scaled 1200
\font\twelvessl=cmssi12
\font\tenssl=cmssi10
\font\eightssl=cmssi8
\font\fourteenssb=cmssbx10 scaled 1440
\font\twelvessb=cmssbx10 scaled 1200
\font\tenssb=cmssbx10
\skewchar\fourteeni='177\skewchar\fourteensy='60
\skewchar\twelvei='177\skewchar\twelvesy='60
\skewchar\ninei='177\skewchar\ninesy='60
\skewchar\eighti='177\skewchar\eightsy='60
\skewchar\sixi='177\skewchar\sixsy='60
\hyphenchar\fourteentt=-1\hyphenchar\twelvett=-1
\hyphenchar\tentt=-1\hyphenchar\ninett=-1
\hyphenchar\eighttt=-1}
%
%
\def\getpcfonts{\message{Installing Almost Computer Modern Fonts.}
\font\fourteenrm=amr10 scaled 1440
\font\twelverm=amr10 scaled 1200
\font\tenrm=amr10
\font\ninerm=amr7 scaled 1200
\font\eightrm=amr7 scaled 1200
\font\sevenrm=amr7
\font\sixrm=amr5 scaled 1200
\font\fiverm=amr5
\font\fourteeni=ammi10 scaled 1440
\font\twelvei=ammi10 scaled 1200
\font\teni=ammi10
\font\ninei=ammi7 scaled 1200
\font\eighti=ammi7 scaled 1200
\font\seveni=ammi7
\font\sixi=ammi5 scaled 1200
\font\fivei=ammi5
\font\fourteensy=amsy10 scaled 1440
\font\twelvesy=amsy10 scaled 1200
\font\tensy=amsy10
\font\ninesy=amsy7 scaled 1200
\font\eightsy=amsy7 scaled 1200
\font\sevensy=amsy7
\font\sixsy=amsy5 scaled 1200
\font\fivesy=amsy5
\font\fourteenex=amex10 scaled 1440
\font\twelveex=amex10 scaled 1200
\font\tenex=amex10
\font\fourteenbf=ambx10 scaled 1440
\font\twelvebf=ambx10 scaled 1200
\font\tenbf=ambx10
\font\ninebf=ambx7 scaled 1200
\font\eightbf=ambx7 scaled 1200
\font\sevenbf=ambx7
\font\sixbf=ambx5 scaled 1200
\font\fivebf=ambx5
\font\fourteenit=amti10 scaled 1440
\font\twelveit=amti10 scaled 1200
\font\tenit=amti10
\let\nineit=\tenit
\let\eightit=\tenit
\let\sevenit=\tenit
\font\fourteensl=amsl10 scaled 1440
\font\twelvesl=amsl10 scaled 1200
\font\tensl=amsl10
\let\ninesl=\tensl
\let\eightsl=\tensl
\font\fourteentt=amtt10 scaled 1440
\font\twelvett=amtt10 scaled 1200
\font\tentt=amtt10
\let\ninett=\tentt
\let\eighttt=\tentt
\let\fourteensc=\fourteenrm        
\let\twelvesc=\twelverm
\let\tensc=\tenrm
\let\fourteenss=\fourteenrm
\let\twelvess=\twelverm
\let\tenss=\tenrm
\let\niness=\ninerm
\let\eightss=\eightrm
\let\fourteenssl=\fourteensl
\let\twelvessl=\twelvesl
\let\tenssl=\tensl
\let\ninessl=\ninesl
\let\eightssl=\eightsl
\let\fourteenssb=\fourteenbf
\let\twelvessb=\twelvebf
\let\tenssb=\tenbf
\skewchar\fourteeni='177\skewchar\fourteensy='60
\skewchar\twelvei='177\skewchar\twelvesy='60
\skewchar\ninei='177\skewchar\ninesy='60
\skewchar\eighti='177\skewchar\eightsy='60
\skewchar\sixi='177\skewchar\sixsy='60
\hyphenchar\fourteentt=-1\hyphenchar\twelvett=-1
\hyphenchar\tentt=-1\hyphenchar\ninett=-1
\hyphenchar\eighttt=-1}
\newfam\scfam
%
%
\def\fourteenpoint{
\def\rm{\fam0\fourteenrm}\def\mit{\fam1\fourteeni}\def\cal{\fam2\fourteensy}
\textfont0=\fourteenrm\scriptfont0=\tenrm\scriptscriptfont0=\sevenrm
\textfont1=\fourteeni\scriptfont1=\teni\scriptscriptfont1=\seveni
\textfont2=\fourteensy\scriptfont2=\tensy\scriptscriptfont2=\sevensy
\textfont3=\fourteenex\scriptfont3=\tenex\scriptscriptfont3=\tenex
\textfont\itfam=\fourteenit\scriptfont\itfam=\tenit
\scriptscriptfont\itfam=\sevenit\def\it{\fam\itfam\fourteenit}
\textfont\slfam=\fourteensl\scriptfont\slfam=\tensl
\scriptscriptfont\slfam=\eightsl\def\sl{\fam\slfam\fourteensl}
\textfont\ttfam=\fourteentt\scriptfont\ttfam=\tentt
\scriptscriptfont\ttfam=\eighttt\def\tt{\fam\ttfam\fourteentt}
\textfont\bffam=\fourteenbf\scriptfont\bffam\tenbf
\scriptscriptfont\bffam=\sevenbf\def\bf{\fam\bffam\fourteenbf}
\textfont\scfam=\fourteensc\scriptfont\scfam=\tensc
\scriptscriptfont\scfam=\tensc\def\sc{\fam\scfam\fourteensc}
\normalbaselineskip=18pt         
\setbox\strutbox=\hbox{\vrule height12.75pt depth5.25pt width0pt}
\normalbaselines\rm}             
\def\twelvepoint{
\def\rm{\fam0\twelverm}\def\mit{\fam1\twelvei}\def\cal{\fam2\twelvesy}
\textfont0=\twelverm\scriptfont0=\eightrm\scriptscriptfont0=\sixrm
\textfont1=\twelvei\scriptfont1=\eighti\scriptscriptfont1=\sixi
\textfont2=\twelvesy\scriptfont2=\eightsy\scriptscriptfont2=\sixsy
\textfont3=\twelveex\scriptfont3=\tenex\scriptscriptfont3=\tenex
\textfont\itfam=\twelveit\scriptfont\itfam=\eightit
\scriptscriptfont\itfam=\sevenit\def\it{\fam\itfam\twelveit}
\textfont\slfam=\twelvesl\scriptfont\slfam=\eightsl
\scriptscriptfont\itfam=\eightsl\def\sl{\fam\slfam\twelvesl}
\textfont\ttfam=\twelvett\scriptfont\ttfam=\eighttt
\scriptscriptfont\ttfam=\eighttt\def\tt{\fam\ttfam\twelvett}
\textfont\bffam=\twelvebf\scriptfont\bffam\eightbf
\scriptscriptfont\bffam=\sixbf\def\bf{\fam\bffam\twelvebf}
\textfont\scfam=\twelvesc\scriptfont\scfam=\tensc
\scriptscriptfont\scfam=\tensc\def\sc{\fam\scfam\twelvesc}
\normalbaselineskip=12pt       
\setbox\strutbox=\hbox{\vrule height8.5pt depth3.5pt width0pt}%
\normalbaselines\rm}           
\def\tenpoint{%
\def\rm{\fam0\tenrm}\def\mit{\fam1\teni}\def\cal{\fam2\tensy}
\textfont0=\tenrm\scriptfont0=\sevenrm\scriptscriptfont0=\fiverm
\textfont1=\teni\scriptfont1=\seveni\scriptscriptfont1=\fivei
\textfont2=\tensy\scriptfont2=\sevensy\scriptscriptfont2=\fivesy
\textfont3=\tenex\scriptfont3=\tenex\scriptscriptfont3=\tenex
\textfont\itfam=\tenit\scriptfont\itfam=\sevenit
\scriptscriptfont\itfam=\sevenit\def\it{\fam\itfam\tenit}
\textfont\slfam=\tensl\scriptfont\slfam=\eightsl
\scriptscriptfont\slfam=\eightsl\def\sl{\fam\slfam\tensl}
\textfont\ttfam=\tentt\scriptfont\ttfam=\eighttt
\scriptscriptfont\ttfam=\eighttt\def\tt{\fam\ttfam\tentt}
\textfont\bffam=\tenbf\scriptfont\bffam\sevenbf
\scriptscriptfont\bffam=\fivebf\def\bf{\fam\bffam\tenbf}
\textfont\scfam=\tensc\scriptfont\scfam=\tensc
\scriptscriptfont\scfam=\tensc\def\sc{\fam\scfam\tensc}
\normalbaselineskip=12pt
\setbox\strutbox=\hbox{\vrule height8.5pt depth3.5pt width0pt}%
\normalbaselines\rm}          
\def\ninepoint{
\def\rm{\fam0\ninerm}\def\mit{\fam1\ninei}\def\cal{\fam2\ninesy}
\textfont0=\ninerm\scriptfont0=\sixrm\scriptscriptfont0=\fiverm
\textfont1=\ninei\scriptfont1=\sixni\scriptscriptfont1=\fivei
\textfont2=\ninesy\scriptfont2=\sixsy\scriptscriptfont2=\fivesy
\textfont3=\tenex\scriptfont3=\tenex\scriptscriptfont3=\tenex
\textfont\itfam=\nineit\scriptfont\itfam=\sevenit
\scriptscriptfont\itfam=\sevenit\def\it{\fam\itfam\nineit}
\textfont\slfam=\ninesl\scriptfont\slfam=\eightsl
\scriptscriptfont\slfam=\eightsl\def\sl{\fam\slfam\ninesl}
\textfont\ttfam=\ninett\scriptfont\ttfam=\eighttt
\scriptscriptfont\ttfam=\eighttt\def\tt{\fam\ttfam\ninett}
\textfont\bffam=\ninebf\scriptfont\bffam\sixbf
\scriptscriptfont\bffam=\fivebf\def\bf{\fam\bffam\ninebf}
\textfont\scfam=\tensc\scriptfont\scfam=\tensc
\scriptscriptfont\scfam=\tensc\def\sc{\fam\scfam\tensc}
\normalbaselineskip=11pt       
\setbox\strutbox=\hbox{\vrule height8pt depth3pt width0pt}%
\normalbaselines\rm}           
%
%
%
%
%
%

%

%
\immediate\write16{The defaults are twelvepoint text, double-spaced, with
no headline, page number footlines and no table of contents. }
\getvaxfonts
 \let\chapfont=\fourteenbf
 \let\secfont=\twelvebf
 \let\subsecfont=\twelvebf
 \let\rightheadfont=\twelveit
 \let\leftheadfont=\twelveit
 \let\centerheadfont=\twelverm
 \let\footfont=\tenrm
 \let\tocfont=\twelvess
%
\def\chapchar{ }    
\def\yesheadlines{\headline={\ifnum\pageno=-1{\hfil} \else \firstmark \fi}}

\def\nofootlines{\footline={\hfil}}
\def\rheader{{\rightheadfont\hfil \chaptype \chapchar}\qquad{\bf\folio}}
\def\lheader{{\bf\folio}\qquad{\leftheadfont \chapname \hfil}}

    \mark{\noexpand\ifodd \noexpand\pageno
                        \noexpand\lheader
        \noexpand\else
                        \noexpand\rheader
        \noexpand\fi}
\newskip\hugeskipamount\hugeskipamount=24pt plus8pt minus8pt
\def\hugeskip{\vskip\hugeskipamount}
\def\hugebreak{\par \ifdim\lastskip<\hugeskipamount\removelastskip
    \penalty-400\hugeskip\fi}
\def\singlespace{\baselineskip=\normalbaselineskip
 \def\secbreak{\medbreak} \def\subsecbreak{\smallbreak}}
\def\oneandhalfspace{\baselineskip=1.5\normalbaselineskip
 \def\secbreak{\bigbreak} \def\subsecbreak{\medbreak}}
\def\doublespace{\baselineskip=2\normalbaselineskip
 \def\secbreak{\hugebreak} \def\subsecbreak{\bigbreak}}
\def\ifthen#1#2#3#4{\if#1#2#3\else#4\fi}
%
%
%

%
%
%
\newcount\chapnum	\chapnum=1
\newcount\secnum	\secnum=0
\newcount\subsecnum	\subsecnum=0
\newcount\eqnum		\eqnum=0
\newcount\fignum	\fignum=0
\newcount\notenum	\notenum=0
\newcount\tabnum	\tabnum=0
\newcount\pgnum	\pgnum=0
\newcount\Items		\Items=0
\newcount\Itemitems		\Itemitems=0
\newcount\Itema		\Itema=0
\newdimen \tempdima  \newdimen \headsize
\newtoks  \temptoka
\newwrite\tocout \newwrite\tofout \newwrite\totout
\def\chapchar{ }    
\def\chaptype{ }  \def\chapname{ }
%
%

\def\eqnumber{{\global\advance\eqnum by 1}  
    \ifthen{\chapchar}{ }                  
      {\gdef\totaleqn{\the\eqnum}}           %
      {\gdef\totaleqn{\chapchar.\the\eqnum}} %
      \eqno{\rm(\totaleqn)}}                
\def\eqnumbera#1{                           
    \ifthen{#1}{a}                          
       {\global\advance\eqnum by 1}         
       {\relax}                             
    \ifthen{\chapchar}{ }                  
      {\gdef\totaleqn{\the\eqnum#1}}         %
      {\gdef\totaleqn{\chapchar.\the\eqnum#1}}
    \eqno{\rm(\totaleqn)}                   %
   }                                        %
\def\eqref#1{                               %
    {\advance \eqnum by -#1                 
    \advance\eqnum by 1}\totaleqn           
    {\advance\eqnum by#1\advance\eqnum by-1}%
   }                                        %
%
\def\fignumber
 {{{\global\advance\fignum by 1             
    }\ifthen{\chapchar}{ }                   
      {\gdef\totalfig{\the\fignum}}         %
      {\gdef\totalfig{\chapchar.\the\fignum}}
    }\totalfig                              %
   }                                        %
\def\fignumbera#1{\ignorespaces             
   \ifthen{#1}{a}                          
      {\global\advance\fignum by 1}         
      {\relax}                             
     \ifthen{#1}{ }                          
      {\global\advance\fignum by 1}         
      {\relax}                             
     \ifthen{\chapchar}{ }                   %
      {\gdef\totalfig{\the\fignum#1}}       %
      {\gdef\totalfig{\chapchar.\the\fignum#1}}
    \totalfig                              %
   }                                        %
\def\figref#1{{                             %
    \advance\fignum by -#1                  
    \advance\fignum by 1}\totalfig{         
    \advance\fignum by #1                   %
     \advance\fignum by-1}                  %
   }                                        %
%
\def\tabnumber{{{\global\advance\tabnum by 1}
    \ifthen{\chapchar}{ }                   
      {\gdef\totaltab{\the\tabnum}}         %
      {\gdef\totaltab{\chapchar.\the\tabnum}}
    }\totaltab                              %
   }                                        %
\def\tabnumbera#1{{                         
    \ifthen{#1}{a}                          
      {\global\advance\tabnum by 1}         
      {\relax}                              
    \ifthen{#1}{ }                          
      {\global\advance\tabnum by 1}         
      {\relax}                              
    \ifthen{\chapchar}{ }                   %
      {\gdef\totaltab{\the\tabnum#1}}       %
      {\gdef\totaltab{\chapchar.\the\tabnum#1}}
    }\totaltab                              %
   }                                        %
\def\tabref#1{                              
    {\advance\tabnum by -#1                 
    \advance\tabnum by 1}\the\tabnum        %
    {\advance\tabnum by #1                  %
    \advance\tabnum by -1}                  %
   }                                        %
\def\tabnam#1{\xdef#1{\the\tabnum}}         
%
\def\note#1{                                
    {\global\advance\notenum by 1}          
    \footnote{$^{\the\notenum}$}            %
    {\footfont#1}                           %
   }                                        %
\def\foot#1{                                
    {\global\advance\notenum by 1}\raise3pt 
    \hbox{\eightrm\the\notenum}\hfil\par    %
    \vskip3pt\hrule\vskip6pt                %
    \noindent\raise3pt                      %
    \hbox{\eightrm \the\notenum}            %
    {\footfont#1\par}                       %
    \vskip6pt\hrule\vskip3pt\noindent       %
   }                                        %
%
\def\figcap#1#2#3
  {{\rm FIGURE\fignumbera{#1}:\ \ #3}
    \immediate\write16{Doing Figure \totalfig}
    \ifthen{\tabletest}{y}                  
      {\ifthen{\chapchar}{ }
        {\def\totaltitle{\the\fignum#1}}
        {\def\totaltitle{\chapchar.\the\fignum#1}}
       \writecontline{\tofout}{1.5em}{2.3em}{\totaltitle}{#2}{.}}
      {\relax}
     }
\def\tabcap#1#2
   {\centerline{\bf TABLE\tabnumbera{#1}}
    \immediate\write16{Doing Table \totaltab}
    \ifthen{\tabletest}{y}                  
      {\ifthen{\chapchar}{ }
        {\def\totaltitle{\the\tabnum#1}}
        {\def\totaltitle{\chapchar.\the\tabnum#1}}
       \writecontline{\totout}{1.5em}{2.3em}{\totaltitle}{#2}{.}}
      {\relax}
     }
\def\chaphead#1#2{
    \vfill\supereject
    \global\advance\chapnum by 1
    \gdef\chapchar{\the\chapnum}
    \secnum=0\subsecnum=0\tabnum=0\fignum=0\eqnum=0
    \immediate\write16{Doing Chapter \chapchar}     
    \centerline{\chapfont PAPER \chapchar}        
    \vskip1.5\normalbaselineskip
    \centerline{\chapfont #1}
    \vskip1.8\normalbaselineskip
    \gdef\chaptype{Paper} \gdef\chapname{ }
    \ifthen{\tabletest}{y}
       {\writecontline{\tocout}{0.0em}{1.5em}{\chapchar}{#1}{\ }}
       {\relax}
    \mark{\noexpand\ifodd \noexpand\pageno
                        \noexpand\lheader
        \noexpand\else
                        \noexpand\rheader
        \noexpand\fi}
    }
\def\apphead#1#2#3{
    \vfill\supereject
    \gdef\chapchar{#1}
    \secnum=0\subsecnum=0\tabnum=0\fignum=0\eqnum=0
    \immediate\write16{Doing Appendix \chapchar}    
    \centerline{\chapfont Appendix \chapchar}       
    \vskip1.5\normalbaselineskip
    \centerline{\chapfont #2}
    \vskip1.8\normalbaselineskip
    \gdef\chaptype{Appendix} \gdef\chapname{#3}
    \ifthen{\tabletest}{y}
       {\writecontline{\tocout}{0.0em}{1.5em}{\chapchar}{#2}{\ }}
       {\relax}
    \mark{\noexpand\ifodd \noexpand\pageno
                        \noexpand\lheader
        \noexpand\else
                        \noexpand\rheader
        \noexpand\fi}
    }
\def\sechead#1{
    \secbreak
    \advance\secnum by 1
    \subsecnum=0
    \ifthen{\chapchar}{ }                      
      {\def\totaltitle{}}
      {\def\totaltitle{\chapchar.\the\secnum}}
    \ifthen{\tabletest}{y}
       {\writecontline{\tocout}{1.5em}{2.3em}{\totaltitle}{#1}{.}}
       {\relax}
    \vskip1.8\normalbaselineskip\par
    \leftline{\secfont\totaltitle#1}
    \nobreak\vskip1.2\normalbaselineskip}
\def\subsechead#1{
    \subsecbreak\global\advance\subsecnum by 1
    \ifthen{\chapchar}{ }                      
      {\def\totaltitle{}}
      {\def\totaltitle{\chapchar.\the\secnum.\the\subsecnum}}
    \ifthen{\tabletest}{y}
       {\writecontline{\tocout}{2.3em}{2.3em}{\totaltitle}{#1}{.}}
       {\relax}
    \vskip1.2\normalbaselineskip\par
    \leftline{\subsecfont\totaltitle\hskip1.5em #1}
    \nobreak\vskip1.0\normalbaselineskip}
%
%
%
\def\Sechead#1{
    \secbreak
    \advance\secnum by 1
    \subsecnum=0
    \ifthen{\chapchar}{ }                      
      {\def\totaltitle{}}
      {\def\totaltitle{\chapchar.\the\secnum}}
    \ifthen{\tabletest}{y}
       {\writecontline{\tocout}{1.5em}{2.3em}{\totaltitle}{#1}{.}}
       {\relax}
    \vskip1.8\normalbaselineskip\par
    {\raggedcenter{\secfont\totaltitle #1}\par}
    \nobreak\vskip1.2\normalbaselineskip}
\def\Subsechead#1{
    \subsecbreak\global\advance\subsecnum by 1
    \ifthen{\chapchar}{ }                      
      {\def\totaltitle{}}
      {\def\totaltitle{\chapchar.\the\secnum.\the\subsecnum}}
    \ifthen{\tabletest}{y}
       {\writecontline{\tocout}{2.3em}{2.3em}{\totaltitle}{#1}{.}}
       {\relax}
    \vskip1.2\normalbaselineskip\par
    {\raggedcenter{\subsecfont\totaltitle #1}\par}
    \nobreak\vskip1.0\normalbaselineskip}
\def\Ssubsechead#1{
    \subsecbreak\global\advance\subsecnum by 1
    \ifthen{\chapchar}{ }                      
      {\def\totaltitle{}}
      {\def\totaltitle{\chapchar.\the\secnum.\the\subsecnum}}
    \ifthen{\tabletest}{y}
       {\writecontline{\tocout}{2.3em}{2.3em}{\totaltitle}{#1}{.}}
       {\relax}
    \vskip1.2\normalbaselineskip\par
    {\raggedcenter{\twelvebf\totaltitle\hskip1.5em #1}\par}
    \nobreak\vskip1.0\normalbaselineskip}
%
%
\def\notables{\def\tabletest{n}}
\def\yestables{\def\tabletest{y}
    \immediate\openout\tocout=\jobname.toc
    \immediate\openout\tofout=\jobname.tof
    \immediate\openout\totout=\jobname.tot\relax}
\def\protect#1{\string#1}
\def\contentsline{\typetocline}
\def\unprotect#1{\csname #1 \endcsname}
\def\writecontline#1#2#3#4#5#6{\begingroup\temptoka={\the\pageno}
    \edef\temp{\write #1{\protect\contentsline{#2}{#3}
    {\protect\numberline{#4}{#5}}{#6}{\the\temptoka}}}
    \temp\endgroup}
%
\def\typetocline#1#2#3#4#5{\vskip0pt plus .2pt \tocfont
   {\hangindent #1\relax\rightskip 2.55em\parfillskip -\rightskip\parindent #1
    \relax\interlinepenalty 10000\leavevmode \tempdima = #2 \relax #3\nobreak
    \leaders\hbox{$\mathsurround=0pt\mkern 4.5mu #4 \mkern 4.5mu$}
    \hfill\nobreak\hbox to 1.55em{\hfil \rm #5}\par}}
\def\numberline#1{\advance\hangindent\tempdima\hbox to \tempdima{#1\hfil}}

\def\bib{\noindent\parskip=4pt\hangindent=2pc\hangafter=1}
\def\rpaper#1#2#3#4#5#6{\bib {#1}\ \ #2.\ \ #3\ {\it\ #4} {\bf\ #5},\ #6.\par}

\def\aa #1#2#3#4#5{\rpaper{#1}{#2}{#3}{Astron.\ Astrophys. }{#4}{#5}}

\def\aj#1#2#3#4#5{\rpaper{#1}{#2}{#3}{Astron.\ J.\/}{#4}{#5}}

\def\apj #1#2#3#4#5{\rpaper{#1}{#2}{#3}{Astrophys.\ J.\/}{#4}{#5}}

\def\baas #1#2#3#4#5{\rpaper{#1}{#2}{#3}{Bull.\ Am.\ Astron.\ Soc.\/}{#4}{#5}}

\def\mnras #1#2#3#4#5{\rpaper{#1}{#2}{#3}{Mon.\ Not.\ Royal Astron.\
Soc.\/}{#4}{#5}}

%


%

%
%

%
%
%

%

%
%
%


%
%

%
%
%


%
\def\asec{\ifmmode^{\prime\prime}\else$\null^{\prime\prime}$\fi}
\def\center#1{\centerline{#1}}
\def\deg{\ifmmode^\circ\else$\null^\circ$\fi}
\def\etal{{\it et al.\/}}


\def\section{\S}

%
%
\newskip\saveparindent
\def\raggedcenter{\leftskip=0pt plus4em\rightskip=\leftskip%
\parfillskip=0pt\spaceskip=.3333em\xspaceskip=.5em%
\pretolerance=9999\tolerance=9999%
\hyphenpenalty=9999\exhyphenpenalty=9999%
\hbadness=9999\saveparindent=\parindent\parindent=0pt}
%
%

%
%
%
\def\twocolumn#1{
     \countdef\inscount=19
   \def\allc##1##2##3##4##5{\global\advance\count1##1 by 1
       \check##1##4##2 \allocationnumber=\count1##1
       \global##3##5=\allocationnumber
       \wlog{\string##5=\string##2\the\allocationnumber}}
   \def\check##1##2##3{\ifnum\count1##1<##2%
         \else\errmessage{No room for a new ##3}\fi}
       \allc{1}{\dimen}{\dimendef}{\inscount}{\fullhsize}
           \fullhsize=6.3 truein
           \hsize=3.1 truein
           \baselineskip=12pt
\def\fullline{\hbox to\fullhsize}
\def\makeheadline{\vbox to 0pt{\vskip-22.5pt
              \fullline{\vbox to 8.5pt{}\the\headline}\vss}
              \nointerlineskip}
\def\makefootline{\baselineskip=24pt \fullline{\the\footline}}
\let\lr=L \allc{4}{\box}{\chardef}{\inscount}{\leftcolumn}
\output={\if L\lr
         \global\setbox\leftcolumn=\columnbox \global\let\lr=R
       \else \doubleformat \global\let\lr=L\fi
       \ifnum\outputpenalty>-20000 \else\dosupereject\fi}
   \def\doubleformat{\shipout\vbox{\makeheadline
      \fullline{\box\leftcolumn\hfil\columnbox}
      \makefootline}
      \advancepageno}
     \def\columnbox{\leftline{\pagebody}}
   { #1 }
\vfill
\if L\lr \else\null\vfill\eject\fi
}
\def\itemperiod{{\global\advance\Items by 1}  %
    \ifthen{\chapchar}{ }                  %
      {\gdef\totalitems{\the\Items}}           %
      {\gdef\totalitems{\chapchar.\the\Items}} %
      \item{\rm\totalitems.}}                %
\def\iitemperiod{{\global\advance\Itemitems by 1}  %
    \ifthen{\chapchar}{ }                  %
      {\gdef\totalitemitems{\the\Itemitems}}           %
      {\gdef\totalitemitems{\chapchar.\the\Itemitems}} %
      \itemitem{\rm\totalitemitems.}}                %
\def\itemperparen{{\global\advance\Items by 1}  %
    \ifthen{\chapchar}{ }                  %
      {\gdef\totalitems{\the\Items}}           %
      {\gdef\totalitems{\chapchar.\the\Items}} %
      \item{\rm\totalitems.)}}                %
\def\iitemperparen{{\global\advance\Itemitems by 1}  %
    \ifthen{\chapchar}{ }                  %
      {\gdef\totalitemitems{\the\Itemitems}}           %
      {\gdef\totalitemitems{\chapchar.\the\Itemitems}} %
      \itemitem{\rm\totalitemitems.)}}                %
\def\itemparen{{\global\advance\Items by 1}  %
    \ifthen{\chapchar}{ }                  %
      {\gdef\totalitems{\the\Items}}           %
      {\gdef\totalitems{\chapchar.\the\Items}} %
      \item{\rm\totalitems)}}                %
\def\iitemparen{{\global\advance\Itemitems by 1}  %
    \ifthen{\chapchar}{ }                  %
      {\gdef\totalitemitems{\the\Itemitems}}           %
      {\gdef\totalitemitems{\chapchar.\the\Itemitems}} %
      \itemitem{\rm\totalitemitems)}}                %
\def\itembullet{{\global\advance\Items by 1}  %
    \ifthen{\chapchar}{ }                  %
      {\gdef\totalitems{\the\Items}}           %
      {\gdef\totalitems{\chapchar.\the\Items}} %
      \item{$\bullet$}}                %

\def\iitembullet{{\global\advance\Itemitems by 1}  %
    \ifthen{\chapchar}{ }                  %
      {\gdef\totalitemitems{\the\Itemitems}}           %
      {\gdef\totalitemitems{\chapchar.\the\Itemitems}} %
      \itemitem{$\bullet$}}                %
\def\itema#1{             %
    \ifthen{#1}{a}       %
       {\global\advance\Items by 1}  %
       {\relax}         %
    \ifthen{\chapchar}{ }                  %
      {\gdef\totalitems{\the\items#1}}         %
      {\gdef\totalitems{\chapchar.\the\items#1}}
    \item{\rm\totalitems.}                   %
   }                                        %
%
%
%
%
%
%
\hyphenation{stand-ard mono-deu-ter-ated}
\hyphenation{ap-plied at-mo-sphere at-mo-spher-ic axi-sym-met-ric
axi-sym-met-rize
 axi-sym-met-rized  axi-sym-met-riz-ation Chan-dra-sek-har con-duct chem-i-cal
 con-duc-tion con-duc-tive con-duc-tiv-i-ty con-vect con-vec-tion
 con-vec-tive Gan-y-mede Gan-y-medean Gan-y-medo-therm Gan-y-medo-therms
 en-gi-neer-ing ele-men-tary
 hom-ol-o-gous hy-dro-dy-nam-ic hy-dro-dy-nam-ics hy-dro-mag-ne-tic
 hy-dro-mag-ne-tics litho-sphere litho-spher-ic mag-neto-hy-dro-dy-nam-ic
 mag-neto-hy-dro-dy-nam-ics mea-sure mea-sured mea-sures mea-sure-ment
 mea-sure-ments non-di-men-sion-al non-di-men-sion-al-ize
 non-di-men-sion-al-ized pa-ram-eter pa-ram-eter-ize pa-ram-eter-ized Pasa-dena
 photo-frag-men-ta-tion para-meterize
 ra-dio-gen-ic ra-dio-nu-clide Rey-nolds Smol-u-chow-ski}

%
\raggedbottom
\def\today{\ifcase\month\or January\or February\or March\or April\or May\or
June\or July\or August\or September\or October\or November\or December\fi,
\space\number\year}

\def\ttoday{\number\day\space \ifcase\month\or January\or February\or March\or
April\or May\or June\or July\or August\or September\or October\or November\or
December\fi \space \number\year}
\def\date{{\ttoday}}
\def\boxit#1{\vbox{\hrule\hbox{\vrule\kern3pt
\vbox{\kern3pt#1\kern3pt}\kern3pt\vrule}\hrule}}

\def\newpage{\vfill \eject}
\def\rnote#1.{$^{\hbox{#1}}$}
\hsize=6.0truein
\vsize=9.0truein
\hoffset=0.25truein
\voffset=0.0truein
\topskip=20pt
\nofootlines
\parindent=20pt
\twelvepoint  
\doublespace  
\yesheadlines
\notables     
\headline={\line{\centerheadfont\hss\ifnum\pageno<2\
\else\folio\fi}}

\newskip\BIGskipamount
\newskip\MEDskipamount
\newskip\SMALLskipamount
\newskip\footbaseskip

\font\titlefont=cmbx10 scaled \magstep2      
\font\secfont=cmbx10 scaled \magstep1        
\font\subfont=cmsl9 scaled  \magstep1        
\font\subsubfont=cmr8 scaled \magstep1


\font\elevenrm=cmr10 scaled \magstephalf  \font\qelevenrm=cmr9
\font\qqelevenrm=cmr7
\font\eleveni=cmmi10 scaled \magstephalf  \font\qeleveni=cmmi9
\font\qqeleveni=cmmi7
\font\elevensy=cmsy10 scaled \magstephalf \font\qelevensy=cmsy9
\font\qqelevensy=cmsy7
\font\elevenbf=cmbx10 scaled \magstephalf \font\qelevenbf=cmbx9
\font\qqelevenbf=cmbx7
\font\elevenit=cmti10 scaled \magstephalf 
\font\elevensl=cmsl10 scaled \magstephalf 
\font\elevenex=cmex10 scaled \magstephalf
\font\elevensc=cmcsc10 scaled \magstephalf

\font\tenrm=cmr10  \font\qtenrm=cmr9 scaled 900  \font\qqtenrm=cmr7 scaled 900
\font\teni=cmmi10  \font\qteni=cmmi9 scaled 900  \font\qqteni=cmmi7 scaled 900
\font\tensy=cmsy10 \font\qtensy=cmsy9 scaled 900 \font\qqtensy=cmsy7 scaled 900
\font\tenbf=cmbx10 \font\qtenbf=cmbx9 scaled 900 \font\qqtenbf=cmbx7 scaled 900
\font\tenit=cmti10 
\font\tensl=cmsl10 \font\qtensl=cmsl9 scaled 900
\font\tenex=cmex10
\font\tensc=cmcsc10

\skewchar\eleveni='177 \skewchar\qeleveni='177 \skewchar\qqeleveni='177
\skewchar\elevensy='60 \skewchar\qelevensy='60 \skewchar\qqelevensy='60
\skewchar\teni='177 \skewchar\qteni='177 \skewchar\qqteni='177
\skewchar\tensy='60 \skewchar\qtensy='60 \skewchar\qqtensy='60


\def\elevenpoint{\def\rm{\fam0\elevenrm}
     \textfont0=\elevenrm \scriptfont0=\qelevenrm
\scriptscriptfont0=\qqelevenrm
     \textfont1=\eleveni  \scriptfont1=\qeleveni  \scriptscriptfont1=\qqeleveni
     \textfont2=\elevensy \scriptfont2=\qelevensy
\scriptscriptfont2=\qqelevensy
     \textfont3=\elevenex \scriptfont3=\elevenex  \scriptscriptfont3=\elevenex
     \textfont4=\eleveni  \scriptfont4=\qeleveni  \scriptscriptfont4=\qqeleveni
     \textfont\itfam=\elevenit \def\it{\fam\itfam\elevenit}%
     \textfont\slfam=\elevensl \def\sl{\fam\slfam\elevensl}%
     \textfont\bffam=\elevenbf \scriptfont\bffam=\qelevenbf
     \scriptscriptfont\bffam=\qqelevenbf \def\bf{\fam\bffam\elevenbf}%
     \normalbaselineskip=13pt
     \setbox\strutbox=\hbox{\vrule height10pt depth4.5pt width0pt}%
      \let\sc=\elevensc \let\smfont=\qelevenrm \normalbaselines\rm}

\def\tenpoint{\def\rm{\fam0\tenrm}
     \textfont0=\tenrm \scriptfont0=\qtenrm \scriptscriptfont0=\qqtenrm
     \textfont1=\teni  \scriptfont1=\qteni  \scriptscriptfont1=\qqteni
     \textfont2=\tensy \scriptfont2=\qtensy \scriptscriptfont2=\qqtensy
     \textfont3=\tenex \scriptfont3=\tenex  \scriptscriptfont3=\tenex
     \textfont\itfam=\tenit \def\it{\fam\itfam\tenit}%
     \textfont\slfam=\tensl \def\sl{\fam\slfam\tensl}%
     \textfont\bffam=\tenbf \scriptfont\bffam=\qtenbf
      \scriptscriptfont\bffam=\qqtenbf \def\bf{\fam\bffam\tenbf}%
     \normalbaselineskip=12pt
     \setbox\strutbox=\hbox{\vrule height8.5pt depth3.5pt width0pt}%
     \let\sc=\tensc \let\smfont=\qtenrm \normalbaselines\rm}

\def\NGC{NGC\kern.33em}
\def\3C{3C\kern.33em}
\def\IC{IC\kern.33em}
\def\3C{3C\kern.33em}
\def\M{M\kern.06em}

\def\puncspace{\ifmmode\,\else{\ifcat.\C{\if.\C\else\if,\C\else\if?\C\else%
\if:\C\else\if;\C\else\if-\C\else\if)\C\else\if/\C\else\if]\C\else\if'\C%
\else\space\fi\fi\fi\fi\fi\fi\fi\fi\fi\fi}%
\else\if\empty\C\else\if\space\C\else\space\fi\fi\fi}\fi}
\def\SP{\let\\=\empty\futurelet\C\puncspace}
%
\def\I.{\kern.2em{\smfont I}}  \def\II.{\kern.2em{\smfont II}}
\def\III.{\kern.2em{\smfont III}} \def\IV.{\kern.2em{\smfont IV}}

\def\eg.{{\it e.g.},}
\def\ie.{{\it i.e.}}
\def\cf.{{\it cf}}
\def\etal.{{\it et al.}}
\def\apj.#1{{\it Ap. J.}, {\bf #1}}
\def\apjl.#1{{\it Ap. J. (Letters)}, {\bf #1}}
\def\apjs.#1{{\it Ap. J. Suppl.}, {\bf #1}}
\def\aj.#1{{\it Astron. J.}, {\bf #1}}
\def\aa.#1{{\it Astr. Ap.}, {\bf #1}}
\def\anrv.#1{{\it Ann. Rev. Astr. Ap.}, {\bf #1}}
\def\mnras.#1{{\it M. N. R. A. S.}, {\bf #1}}
\def\baas.#1{{\it Bull. A. A. S.}, {\bf #1}}
\def\pasp.#1{{\it P. A. S. P.}, {\bf #1}}

\def\kms.{km~s$^{-1}$}
\def\mic.{$\mu$m}
\def\um.{$\mu$m}
\def\msun.{$M_\odot$}
\def\lsun.{$L_\odot$}
\def\13CO.{$^{13}$CO}
\def\C18O.{C$^{18}$O}

\def\degree#1{\ifmmode{\if.#1{{^\circ}\llap.}\else{^\circ} #1\fi}\else
{\if.#1$^\circ$\llap.\else\if\empty#1$^\circ$#1\else$^\circ$ #1\fi\fi}\fi}
\def\arcmin#1{\ifmmode{\if.#1{'\llap.}\else{'} #1\fi}\else
{\if.#1$'$\llap.\else$'$ #1\fi}\fi}
\def\arcsec#1{\ifmmode{\if.#1{''\llap.}\else{''} #1\fi}\else
{\if.#1$''$\llap.\else$''$ #1\fi}\fi}
\def\sun{\ifmmode _\odot \else $_{\odot}$\fi\SP}
\def\earth{\ifmmode _\oplus \else $_{\oplus}$\fi\SP}

%


 \mathcode`*="002A 
\def\simless{\mathbin{\lower 3pt\hbox
     {$\rlap{\raise 5pt\hbox{$\char'074$}}\mathchar"7218$}}} 
\def\simgreat{\mathbin{\lower 3pt\hbox
     {$\rlap{\raise 5pt\hbox{$\char'076$}}\mathchar"7218$}}} 
\def\about{{$\sim$}}

\def\e#1{\eqno{(#1)}}

\def\boxit#1{\vbox{\hrule\hbox{\vrule\hskip 3pt \vbox{\vskip 3pt \hbox{#1}
     \vskip 3pt}\hskip 3pt\vrule}\hrule}}

\def\msg#1{{\def\\{}\message{#1}}}
  
\def\center#1{{\def\\{\par}\rightskip=0pt plus1fil \leftskip=\rightskip
     \parindent=0pt \parfillskip=0pt #1\par}}
\def\newline{\hfil\break}
\def\newpage{\vfill\eject}
\def\SMALLskip{\vglue\SMALLskipamount}
\def\MEDskip{\vglue\MEDskipamount}
\def\BIGskip{\vglue\BIGskipamount}

\def\nopagenumber{\global\nopagenumbertrue}
\newif\ifnopagenumber \nopagenumberfalse
\footline={}
\headline={\ifnopagenumber {} \global\nopagenumberfalse
     \else{\hss\elevenrm-- \folio\ --\hss}\fi}

\catcode`\@=11
\def\vfootnote#1{\insert\footins\bgroup\tenpoint\baselineskip12pt
     \interlinepenalty\interfootnotelinepenalty
      \splittopskip\ht\strutbox 
       \splitmaxdepth\dp\strutbox \floatingpenalty=20000
        \parfillskip=0pt plus1fil \parindent=25pt
         \leftskip=0pt \rightskip=0pt \spaceskip=0pt \xspaceskip=0pt
          \textindent{#1}\footstrut\futurelet\next\fo@t}
\catcode`\@=12
\newcount\notenumber

\def\note{\global\advance\notenumber by 1 \footnote{$^{\the\notenumber}$}}

\def\titlepage{\hrule height 0pt \nobreak\vskip 0pt plus0.8fil
     \global\nopagenumbertrue \baselineskip20pt}
\def\endtitlepage{\vskip 0pt plus 1.3fil \normalbaselines \eject}
\def\title#1{\vskip 0pt plus .1\vsize
     \center{\baselineskip 20pt \BIGskip \let\smfont=\secfont \titlefont\ #1}}
\def\author#1{\center{\baselineskip 20pt \MEDskip \sc\ #1}}
\def\affiliation#1{\center{\baselineskip 20pt \vskip0pt \elevenrm\ #1}}
\def\date#1{\center{\baselineskip 20pt \BIGskip \it\ #1}}
\def\and{\baselineskip 20pt \MEDskip \center{\sc\ and}}


\def\abstract{\BIGskip \begingroup \centerline{ABSTRACT} \leftskip=20pt
     \rightskip=20pt \baselineskip=14pt \parindent=15pt \nobreak\SMALLskip
      \parskip=0pt}


\def\section#1\par{\vskip0pt plus.05\vsize\penalty-100
     \BIGskip \vskip\parskip
      \msg{#1}\center{\baselineskip 15pt \let\smfont=\tenbf
       \secfont\ #1}\nobreak\MEDskip}
\def\subsection#1\par{\vskip0pt plus.03\vsize\penalty-60 \MEDskip\vskip\parskip
     \msg{#1}\center{\baselineskip 15pt \let\smfont=\tensl
      \subfont\ #1}\nobreak\MEDskip}
\def\subsubsection#1\par{\vskip0pt plus.02\vsize\penalty-40 \SMALLskip\vskip
     \parskip\msg{#1}\center{\baselineskip 15pt \let\smfont=\qtensl
      \subsubfont\ #1}\nobreak\SMALLskip}



\def\bysame{\hbox to 80pt{\leaders\hrule height 2.4pt depth -2pt\hfill .\ }}

\def\bysame{\hbox to 80pt{\leaders\hrule height 2.4pt depth -2pt\hfill .\ }}


\def\figure#1{\par\hangindent40pt \noindent{Figure #1: }\nobreak}

 \def\chaphead{}

\newcount\tablenumber \newbox\tabbox \newdimen\tabwid
\def\clrtablenumber{\global\tablenumber=0} \clrtablenumber
\def\lasttable{{\rm\chaphead\the\tablenumber}\SP}
\def\nexttable{\advance\tablenumber by1 {\rm\chaphead\the\tablenumber}%
   \advance\tablenumber by-1 \SP}
\def\PM{\ifmmode\pm\else${}\pm{}$\fi}
\def\trule{\noalign{\vskip4pt}\noalign{\hrule}\noalign{\vskip4pt}}

\let\tablefont=\tenpoint
\font\sc=cmcsc10
\newdimen\tindent \newskip\tbaseskip \tbaseskip=12pt

{\catcode`\?=\active \catcode`\"=\active \catcode`\^^M=\active%
\gdef\table#1[#2]{\global\advance\tablenumber by1\begingroup%
  \ifx#1\empty\else\xdef#1{\chaphead\the\tablenumber\noexpand\SP}\fi%
  \tablefont \baselineskip\tbaseskip%
  \def\hhh{\hskip0pt plus1000pt}\def\?{\char`\?}%
  \def\a{\rlap{$^a$}} \def\b{\rlap{$^b$}} \def\c{\rlap{$^c$}}%
  \def\d{\rlap{$^d$}} \def\e{\rlap{$^e$}} \def\f{\rlap{$^f$}}%
  \def\g{\rlap{$^g$}} \def\h{\rlap{$^h$}} \def\i{\rlap{$^i$}}%
  \def\j{\rlap{$^j$}} \def\k{\rlap{$^k$}} \def\l{\rlap{$^l$}}%
  \def\|##1{\rlap{$^{\rm ##1}$}}%
  \def\ms##1{\multispan ##1}%
\def\...{\null\nobreak\leaders\hbox to0.5em{\hss.\hss}\hskip1.5em plus1filll\
}%
  \catcode`\"=\active%
  \catcode`\^^I=4 \catcode`\^^M=\active \catcode`~=\active \catcode`?=\active%
  \def?{\enspace} \def~{\hfil} \def^^M{\crcr} \def"{\quad}\tabskip=0pt%
  \pageinsert\vbox to\vsize\bgroup\hrule height0pt\vfil%
  \centerline{TABLE \chaphead\the\tablenumber}%
  \def\tablename{#2}\if\tablename\empty\else\centerline{{\sc\tablename}}\fi%
  \setbox\tabbox=\vbox\bgroup\halign\bgroup}%
\gdef\tabcont{\begingroup%
  \tablefont \baselineskip\tbaseskip%
  \def\hhh{\hskip0pt plus1000pt}\def\?{\char`\?}%
  \def\a{\rlap{$^a$}} \def\b{\rlap{$^b$}} \def\c{\rlap{$^c$}}%
  \def\d{\rlap{$^d$}} \def\e{\rlap{$^e$}} \def\f{\rlap{$^f$}}%
  \def\g{\rlap{$^g$}} \def\h{\rlap{$^h$}} \def\i{\rlap{$^i$}}%
  \def\j{\rlap{$^j$}} \def\k{\rlap{$^k$}} \def\l{\rlap{$^l$}}%
  \def\|##1{\rlap{$^{\rm ##1}$}}%
  \def\ms##1{\multispan ##1}%
\def\...{\null\nobreak\leaders\hbox to0.5em{\hss.\hss}\hskip1.5em plus1filll\
}%
  \catcode`\"=\active%
  \catcode`\^^I=4 \catcode`\^^M=\active \catcode`~=\active \catcode`?=\active%
  \def?{\enspace} \def~{\hfil} \def^^M{\crcr} \def"{\quad}\tabskip=0pt%
  \pageinsert\vbox to\vsize\bgroup\hrule height0pt\vfil%
  \centerline{TABLE \chaphead\the\tablenumber{ \it (continued)}}%
  \setbox\tabbox=\vbox\bgroup\halign\bgroup}%
\gdef\tnotes{\trule\egroup\egroup\tabwid=\wd\tabbox%
  \hbox to\hsize{\hfill{\box\tabbox}\hfill}%
  \def\a{{$^a$}} \def\b{{$^b$}} \def\c{{$^c$}}%
  \def\d{{$^d$}} \def\e{{$^e$}} \def\f{{$^f$}}%
  \def\g{{$^g$}} \def\h{{$^h$}} \def\i{{$^i$}}%
  \def\j{{$^j$}} \def\k{{$^k$}} \def\l{{$^l$}}%
  \def\|##1{{$^{\rm ##1}$}}%
  \baselineskip\footbaseskip%
  \advance\tabwid by-\hsize\divide\tabwid by-2\advance\tabwid by20pt%
  \leftskip=\tabwid%
  \rightskip=\tabwid \parskip=1pt \parindent=0pt \catcode`\^^M=5%
  \def\par{\endgraf\leftskip=\tabwid\rightskip=\tabwid\hangindent=10pt}%
  \def\endtable{\par\vfil\egroup\endinsert\endgroup}\hangindent=10pt}}%
\def\endtable{\trule\egroup\egroup\hbox to\hsize{\hfill{\box\tabbox}\hfill}
  \vfil\egroup\endinsert\endgroup}

\elevenpoint
\SMALLskipamount=4pt plus1pt
\MEDskipamount=8pt plus2pt
\BIGskipamount=16pt plus4pt
\footbaseskip=16pt
\parindent=30pt
\hsize 15 truecm
\vsize 20.5 truecm
\hoffset 0.8 truecm
\voffset 1.0 truecm
\hrule height0pt
\nopagenumber

\def\lax    {\ifmmode{_<\atop^{\sim}}\else{${_<\atop^{\sim}}$}\fi}
\def\gax    {\ifmmode{_>\atop^{\sim}}\else{${_>\atop^{\sim}}$}\fi}

\def\asec{$^{\prime\prime}$}
\def\deg{$^{\circ}$}

\def\h2{$\rm H_2$}
\def\nh2{n_{H_2}}
\def\2to1{2\!\!\!\!\to\!\!\!\!1}
\def\oneto0{1\!\!\!\!\to\!\!\!\!0}

\def\C{{\fam0 C}}

\def\13CO{{\fam0 ^{13}CO}}

\def\M{{\fam0\,M}}

\def\O1D{{\fam0 O(^1D)}}

\def\O{{\fam0 O}}

\def\S{{\fam0 S}}

\def\arcsec{{\fam0 \, arcsec}}

\def\g{{\fam0\, g}}

\def\scinot#1.{\hbox{$\,$ $\times$ $10^{#1}$}}

\def\ten#1.{\hbox{$10^{#1}$}}
\def\to{\hbox{$\hphantom{1} \rightarrow \hphantom{1}$}}
\def\spose#1{\hbox to 0pt{#1\hss}}
\def\lta{\mathrel{\spose{\lower 3pt\hbox{$\mathchar "218$}}\raise
2.0pt\hbox{$\mathchar"13C$}}}
\def\gta{\mathrel{\spose{\lower 3pt\hbox{$\mathchar "218$}}\raise
2.0pt\hbox{$\mathchar"13E$}}}
\def\lrarrow{\mathrel{\spose{\lower 1pt\hbox{$\rightarrow$}}\raise
3.0pt\hbox{$\leftarrow$}}}
\def\qovertilde{\mathrel{\spose{\lower 4pt\hbox{$q$}}\raise
1.0pt\hbox{$\mathchar "218$}}}
\def\Btilde{\mathrel{\spose{\lower 4pt\hbox{$\mathchar "218$}}\raise
1.0pt\hbox{$B$}}}
\def\Cundertilde{\mathrel{\spose{\lower 4pt\hbox{$\mathchar "218$}}\raise
1.0pt\hbox{$C$}}}
\def\htilde{\mathrel{\spose{\hbox{$h$}}\raise 6.0pt\hbox{$\mathchar "218$}}}
\def\nablatilde{\mathrel{\spose{\lower 4pt\hbox{$\mathchar "218$}}\raise
1.0pt\hbox{$\nabla$}}}
\def\ntilde{\mathrel{\spose{\lower 4pt\hbox{$\mathchar "218$}}\raise
1.0pt\hbox{$n$}}}
\def\omegatilde{\mathrel{\spose{\lower 4pt\hbox{$\mathchar "218$}}\raise
1.0pt\hbox{$\omega$}}}
\def\Omegatilde{\mathrel{\spose{\lower 4pt\hbox{$\mathchar "218$}}\raise
1.0pt\hbox{$\Omega$}}}
\def\omegauptilde{\mathrel{\spose{\hbox{$\omega$}}\raise 6.0pt\hbox{$\mathchar
"218$}}}

\def\phitilde{\mathrel{\spose{\lower 4pt\hbox{$\mathchar "218$}}\raise
1.0pt\hbox{$\phi$}}}
\def\rtilde{\mathrel{\spose{\lower 4pt\hbox{$\mathchar "218$}}\raise
1.0pt\hbox{$r$}}}
\def\stilde{\mathrel{\spose{\lower 4pt\hbox{$\mathchar "218$}}\raise
1.0pt\hbox{$s$}}}
\def\Oundertilde{\mathrel{\spose{\lower 4pt\hbox{$\mathchar "218$}}\raise
1.0pt\hbox{$O$}}}
\def\Sundertilde{\mathrel{\spose{\lower 4pt\hbox{$\mathchar "218$}}\raise
1.0pt\hbox{$S$}}}
\def\gundertilde{\mathrel{\spose{\lower 4pt\hbox{$\mathchar "218$}}\raise
1.0pt\hbox{$g$}}}
\def\thetatilde{\mathrel{\spose{\lower 4pt\hbox{$\mathchar "218$}}\raise
1.0pt\hbox{$\theta$}}}
\def\utilde{\mathrel{\spose{\lower 4pt\hbox{$\mathchar "218$}}\raise
1.0pt\hbox{$u$}}}
\def\vtilde{\mathrel{\spose{\lower 4pt\hbox{$\mathchar "218$}}\raise
1.0pt\hbox{$v$}}}
\def\Vundertilde{\mathrel{\spose{\lower 4pt\hbox{$\mathchar "218$}}\raise
1.0pt\hbox{$V$}}}
\def\ztilde{\mathrel{\spose{\lower 4pt\hbox{$\mathchar "218$}}\raise
1.0pt\hbox{$z$}}}
\def\ang{\spose{\hbox{A}}\raise 2.0pt\hbox{\char'27}}
\def\<{\hbox{$\langle$}}
\def\ff#1.{\hbox{$\null^{#1}$}}
\def\title#1{{#1}}

\def\Pasadena9{{Pasadena, California\ \ \ 91109}}

\def\date{{\ttoday}}

\def\permil{\%\hskip-.1em\relax{\smash{\lower46576sp\hbox{$\scriptstyle 0$}}}}


\def\figure#1{\par\hangindent40pt \noindent{Figure #1: }\nobreak}

\def\eg.{{\it e.g.},}
\def\ie.{{\it i.e.}}
\def\cf.{{\it cf}}
\def\etal.{{\it et al.}}
\def\apj.#1{{\it Ap. J.}, {\bf #1}}
\def\apjl.#1{{\it Ap. J. (Letters)}, {\bf #1}}
\def\apjs.#1{{\it Ap. J. Suppl.}, {\bf #1}}
\def\aj.#1{{\it Astron. J.}, {\bf #1}}
\def\aa.#1{{\it Astr. Ap.}, {\bf #1}}
\def\anrv.#1{{\it Ann. Rev. Astr. Ap.}, {\bf #1}}
\def\mnras.#1{{\it M. N. R. A. S.}, {\bf #1}}
\def\baas.#1{{\it Bull. A. A. S.}, {\bf #1}}
\def\pasp.#1{{\it P. A. S. P.}, {\bf #1}}

\def\kms.{km~s$^{-1}$}
\def\mic.{$\mu$m}
\def\um.{$\mu$m}
\def\msun.{$M_\odot$}
\def\lsun.{$L_\odot$}
\def\13CO.{$^{13}$CO}
\def\C18O.{C$^{18}$O}

\def\degree#1{\ifmmode{\if.#1{{^\circ}\llap.}\else{^\circ} #1\fi}\else
{\if.#1$^\circ$\llap.\else\if\empty#1$^\circ$#1\else$^\circ$ #1\fi\fi}\fi}
\def\arcmin#1{\ifmmode{\if.#1{'\llap.}\else{'} #1\fi}\else
{\if.#1$'$\llap.\else$'$ #1\fi}\fi}
\def\arcsec#1{\ifmmode{\if.#1{''\llap.}\else{''} #1\fi}\else
{\if.#1$''$\llap.\else$''$ #1\fi}\fi}
\def\sun{\ifmmode _\odot \else $_{\odot}$\fi\SP}
\def\earth{\ifmmode _\oplus \else $_{\oplus}$\fi\SP}

%


 \mathcode`*="002A 
\def\simless{\mathbin{\lower 3pt\hbox
     {$\rlap{\raise 5pt\hbox{$\char'074$}}\mathchar"7218$}}} 
\def\simgreat{\mathbin{\lower 3pt\hbox
     {$\rlap{\raise 5pt\hbox{$\char'076$}}\mathchar"7218$}}} 
\def\about{{$\sim$}}


\def\lax    {\ifmmode{_<\atop^{\sim}}\else{${_<\atop^{\sim}}$}\fi}
\def\gax    {\ifmmode{_>\atop^{\sim}}\else{${_>\atop^{\sim}}$}\fi}

\def\asec{$^{\prime\prime}$}
\def\deg{$^{\circ}$}

\def\h2{$\rm H_2$}
\def\nh2{n_{H_2}}

\def\singlespace{\baselineskip 12pt \lineskip 1pt \parskip 2pt plus 1 pt}
\def\mediumspace{\baselineskip 18pt \lineskip 6pt \parskip 3pt plus 5 pt}
\def\mediumspace2{\baselineskip 16pt \lineskip 5pt \parskip 2pt plus 2 pt}
\def\doublespace{\baselineskip 24pt \lineskip 10pt \parskip 5pt plus 10 pt}
\def\today{\number\day\enspace
     \ifcase\month\or January\or Febuary\or March\or April\or May\or
     June\or July\or August\or September\or October\or
     November\or December\fi \enspace\number\year}
\def\clock{\count0=\time \divide\count0 by 60
    \count1=\count0 \multiply\count1 by -60 \advance\count1 by \time
    \number\count0:\ifnum\count1<10{0\number\count1}\else\number\count1\fi}

\def\deg{\ifmmode^\circ\else$^\circ$\fi}

\def\jref#1 #2 #3 #4 {{\par\noindent \hangindent=3em
      \advance \rightskip by 0em #1, {\it#2}, {\bf#3}, #4.\par}}
\def\ref#1{{\par\noindent \hangindent=3em
      \advance \rightskip by 0em #1.\par}}
\newcount\eqnum
\def\nexteq{\global\advance\eqnum by1 \eqno(\number\eqnum)}
\def\lasteq#1{\if)#1[\number\eqnum]\else(\number\eqnum)\fi#1}
\def\preveq#1#2{{\advance\eqnum by-#1
    \if)#2[\number\eqnum]\else(\number\eqnum)\fi}#2}
\def\endtable{\endgroup}
\def\tableheight{\vrule width 0pt height 8.5pt depth 3.5pt}
{\catcode`|=\active \catcode`&=\active
    \gdef\tabledelim{\catcode`|=\active \let|=\vbar
                     \catcode`&=\active \let&=\nobar} }
\def\table{\begingroup
    \def\twidth{\hsize}
    \def\tablewidth##1{\def\twidth{##1}}
    \def\defaultheight{\vrule width 0pt height 8.5pt depth 3.5pt}
    \def\heightdepth##1{\dimen0=##1
        \ifdim\dimen0>5pt
            \divide\dimen0 by 2 \advance\dimen0 by 2.5pt
            \dimen1=\dimen0 \advance\dimen1 by -5pt
            \vrule width 0pt height \the\dimen0  depth \the\dimen1
        \else  \divide\dimen0 by 2
            \vrule width 0pt height \the\dimen0  depth \the\dimen0 \fi}
    \def\spacing##1{\def\defaultheight{\heightdepth{##1}}}
    \def\nextheight##1{\noalign{\gdef\tableheight{\heightdepth{##1}}}}
    \def\end{\cr\noalign{\gdef\tableheight{\defaultheight}}}
    \def\zerowidth##1{\omit\hidewidth ##1 \hidewidth}
    \def\hline{\noalign{\hrule}}
    \def\skip##1{\noalign{\vskip##1}}
    \def\bskip##1{\noalign{\hbox to \twidth{\vrule height##1 depth 0pt \hfil
        \vrule height##1 depth 0pt}}}
    \def\header##1{\noalign{\hbox to \twidth{\hfil ##1 \unskip\hfil}}}
    \def\bheader##1{\noalign{\hbox to \twidth{\vrule\hfil ##1
        \unskip\hfil\vrule}}}
    \def\spanloop{\span\omit \advance\mscount by -1}
    \def\extend##1##2{\omit
        \mscount=##1 \multiply\mscount by 2 \advance\mscount by -1
        \loop\ifnum\mscount>1 \spanloop\repeat \ \hfil ##2 \unskip\hfil}
    \def\vbar{&\vrule&}
    \def\nobar{&&}
    \def\hdash##1{ \noalign{ \relax \gdef\tableheight{\heightdepth{0pt}}
        \toks0={} \count0=1 \count1=0 \putout##1\end
        \toks0=\expandafter{\the\toks0 &\end} \xdef\piggy{\the\toks0} }
        \piggy}
    \let\e=\expandafter
    \def\putspace{\ifnum\count0>1 \advance\count0 by -1
        \toks0=\e\e\e{\the\e\toks0\e&\e\multispan\e{\the\count0}\hfill}
        \fi \count0=0 }
    \def\putrule{\ifnum\count1>0 \advance\count1 by 1

\toks0=\e\e\e{\the\e\toks0\e&\e\multispan\e{\the\count1}\leaders\hrule\hfill}
        \fi \count1=0 }
    \def\putout##1{\ifx##1\end \putspace \putrule \let\next=\relax
        \else \let\next=\putout
            \ifx##1- \advance\count1 by 2 \putspace
            \else    \advance\count0 by 2 \putrule \fi \fi \next}   }
\def\tablespec#1{
    \def\vdimens{\noexpand\tableheight}
    \def\tabby{\tabskip=0pt plus100pt minus100pt}
    \def\r{&################\tabby&\hfil################\unskip}
    \def\c{&################\tabby&\hfil################\unskip\hfil}
    \def\l{&################\tabby&################\unskip\hfil}
    \edef\templ{\noexpand\vdimens ########\unskip  #1
         \unskip&########\tabskip=0pt&########\cr}
    \tabledelim
    \edef\body##1{ \vbox{
        \tabskip=0pt \offinterlineskip
        \halign to \twidth {\templ ##1}}} }

\def\kms.{km~s$^{-1}$}
\def\mic.{$\mu$m}
\def\um.{$\mu$m}
\def\msun.{$M_\odot$}
\def\lsun.{$L_\odot$}
\def\13CO.{$^{13}$CO}
\def\C18O.{C$^{18}$O}

\def\simless{\mathbin{\lower 3pt\hbox
     {$\rlap{\raise 5pt\hbox{$\char'074$}}\mathchar"7218$}}} 
\def\simgreat{\mathbin{\lower 3pt\hbox
     {$\rlap{\raise 5pt\hbox{$\char'076$}}\mathchar"7218$}}} 
\def\about{{$\sim$}}


\def\lax    {\ifmmode{_<\atop^{\sim}}\else{${_<\atop^{\sim}}$}\fi}
\def\gax    {\ifmmode{_>\atop^{\sim}}\else{${_>\atop^{\sim}}$}\fi}

\def\asec{$^{\prime\prime}$}
\def\deg{$^{\circ}$}

\def\h2{$\rm H_2$}
\def\nh2{n_{H_2}}


\def\figure#1{\par\hangindent40pt \noindent{Figure #1: }\nobreak}

\newskip\BIGskipamount
\newskip\MEDskipamount
\newskip\SMALLskipamount
\newskip\footbaseskip

\font\titlefont=cmbx10 scaled \magstep2      
\font\secfont=cmbx10 scaled \magstep1        
\font\subfont=cmsl9 scaled  \magstep1        
\font\subsubfont=cmr8 scaled \magstep1


\font\elevenrm=cmr10 scaled \magstephalf  \font\qelevenrm=cmr9
\font\qqelevenrm=cmr7
\font\eleveni=cmmi10 scaled \magstephalf  \font\qeleveni=cmmi9
\font\qqeleveni=cmmi7
\font\elevensy=cmsy10 scaled \magstephalf \font\qelevensy=cmsy9
\font\qqelevensy=cmsy7
\font\elevenbf=cmbx10 scaled \magstephalf \font\qelevenbf=cmbx9
\font\qqelevenbf=cmbx7
\font\elevenit=cmti10 scaled \magstephalf 
\font\elevensl=cmsl10 scaled \magstephalf 
\font\elevenex=cmex10 scaled \magstephalf
\font\elevensc=cmcsc10 scaled \magstephalf

\font\tenrm=cmr10  \font\qtenrm=cmr9 scaled 900  \font\qqtenrm=cmr7 scaled 900
\font\teni=cmmi10  \font\qteni=cmmi9 scaled 900  \font\qqteni=cmmi7 scaled 900
\font\tensy=cmsy10 \font\qtensy=cmsy9 scaled 900 \font\qqtensy=cmsy7 scaled 900
\font\tenbf=cmbx10 \font\qtenbf=cmbx9 scaled 900 \font\qqtenbf=cmbx7 scaled 900
\font\tenit=cmti10 
\font\tensl=cmsl10 \font\qtensl=cmsl9 scaled 900
\font\tenex=cmex10
\font\tensc=cmcsc10

\skewchar\eleveni='177 \skewchar\qeleveni='177 \skewchar\qqeleveni='177
\skewchar\elevensy='60 \skewchar\qelevensy='60 \skewchar\qqelevensy='60
\skewchar\teni='177 \skewchar\qteni='177 \skewchar\qqteni='177
\skewchar\tensy='60 \skewchar\qtensy='60 \skewchar\qqtensy='60


\def\elevenpoint{\def\rm{\fam0\elevenrm}
     \textfont0=\elevenrm \scriptfont0=\qelevenrm
\scriptscriptfont0=\qqelevenrm
     \textfont1=\eleveni  \scriptfont1=\qeleveni  \scriptscriptfont1=\qqeleveni
     \textfont2=\elevensy \scriptfont2=\qelevensy
\scriptscriptfont2=\qqelevensy
     \textfont3=\elevenex \scriptfont3=\elevenex  \scriptscriptfont3=\elevenex
     \textfont4=\eleveni  \scriptfont4=\qeleveni  \scriptscriptfont4=\qqeleveni
     \textfont\itfam=\elevenit \def\it{\fam\itfam\elevenit}%
     \textfont\slfam=\elevensl \def\sl{\fam\slfam\elevensl}%
     \textfont\bffam=\elevenbf \scriptfont\bffam=\qelevenbf
     \scriptscriptfont\bffam=\qqelevenbf \def\bf{\fam\bffam\elevenbf}%
     \normalbaselineskip=13pt
     \setbox\strutbox=\hbox{\vrule height10pt depth4.5pt width0pt}%
      \let\sc=\elevensc \let\smfont=\qelevenrm \normalbaselines\rm}

\def\tenpoint{\def\rm{\fam0\tenrm}
     \textfont0=\tenrm \scriptfont0=\qtenrm \scriptscriptfont0=\qqtenrm
     \textfont1=\teni  \scriptfont1=\qteni  \scriptscriptfont1=\qqteni
     \textfont2=\tensy \scriptfont2=\qtensy \scriptscriptfont2=\qqtensy
     \textfont3=\tenex \scriptfont3=\tenex  \scriptscriptfont3=\tenex
     \textfont\itfam=\tenit \def\it{\fam\itfam\tenit}%
     \textfont\slfam=\tensl \def\sl{\fam\slfam\tensl}%
     \textfont\bffam=\tenbf \scriptfont\bffam=\qtenbf
      \scriptscriptfont\bffam=\qqtenbf \def\bf{\fam\bffam\tenbf}%
     \normalbaselineskip=12pt
     \setbox\strutbox=\hbox{\vrule height8.5pt depth3.5pt width0pt}%
     \let\sc=\tensc \let\smfont=\qtenrm \normalbaselines\rm}

\def\NGC{NGC\kern.33em}
\def\3C{3C\kern.33em}
\def\IC{IC\kern.33em}
\def\3C{3C\kern.33em}
\def\M{M\kern.06em}

\def\puncspace{\ifmmode\,\else{\ifcat.\C{\if.\C\else\if,\C\else\if?\C\else%
\if:\C\else\if;\C\else\if-\C\else\if)\C\else\if/\C\else\if]\C\else\if'\C%
\else\space\fi\fi\fi\fi\fi\fi\fi\fi\fi\fi}%
\else\if\empty\C\else\if\space\C\else\space\fi\fi\fi}\fi}
\def\SP{\let\\=\empty\futurelet\C\puncspace}
%
\def\I.{\kern.2em{\smfont I}}  \def\II.{\kern.2em{\smfont II}}
\def\III.{\kern.2em{\smfont III}} \def\IV.{\kern.2em{\smfont IV}}

\def\eg.{{\it e.g.},}
\def\ie.{{\it i.e.}}
\def\cf.{{\it cf}}
\def\etal.{{\it et al.}}
\def\apj.{ApJ.}
\def\apjl.#1{{\it Ap. J. (Letters)}, {\bf #1}}
\def\apjs.#1{{\it Ap. J. Suppl.}, {\bf #1}}
\def\aj.#1{{\it Astron. J.}, {\bf #1}}
\def\aa.{A$\&$A}
\def\araa.{ARA$\&$A}
\def\mnras.{MNRAS}
\def\baas.#1{{\it Bull. A. A. S.}, {\bf #1}}
\def\pasp.#1{{\it P. A. S. P.}, {\bf #1}}

\def\kms.{km~s$^{-1}$}
\def\mic.{$\mu$m}
\def\um.{$\mu$m}
\def\msun.{$M_\odot$}
\def\lsun.{$L_\odot$}
\def\13CO.{$^{13}$CO}
\def\C18O.{C$^{18}$O}

\def\degree#1{\ifmmode{\if.#1{{^\circ}\llap.}\else{^\circ} #1\fi}\else
{\if.#1$^\circ$\llap.\else\if\empty#1$^\circ$#1\else$^\circ$ #1\fi\fi}\fi}
\def\arcmin#1{\ifmmode{\if.#1{'\llap.}\else{'} #1\fi}\else
{\if.#1$'$\llap.\else$'$ #1\fi}\fi}
\def\arcsec#1{\ifmmode{\if.#1{''\llap.}\else{''} #1\fi}\else
{\if.#1$''$\llap.\else$''$ #1\fi}\fi}
\def\sun{\ifmmode _\odot \else $_{\odot}$\fi\SP}
\def\earth{\ifmmode _\oplus \else $_{\oplus}$\fi\SP}

%


 \mathcode`*="002A 
\def\simless{\mathbin{\lower 3pt\hbox
     {$\rlap{\raise 5pt\hbox{$\char'074$}}\mathchar"7218$}}} 
\def\simgreat{\mathbin{\lower 3pt\hbox
     {$\rlap{\raise 5pt\hbox{$\char'076$}}\mathchar"7218$}}} 
\def\about{{$\sim$}}

\def\e#1{\eqno{(#1)}}

\def\boxit#1{\vbox{\hrule\hbox{\vrule\hskip 3pt \vbox{\vskip 3pt \hbox{#1}
     \vskip 3pt}\hskip 3pt\vrule}\hrule}}

\def\msg#1{{\def\\{}\message{#1}}}
  
\def\center#1{{\def\\{\par}\rightskip=0pt plus1fil \leftskip=\rightskip
     \parindent=0pt \parfillskip=0pt #1\par}}
\def\newline{\hfil\break}
\def\newpage{\vfill\eject}
\def\SMALLskip{\vglue\SMALLskipamount}
\def\MEDskip{\vglue\MEDskipamount}
\def\BIGskip{\vglue\BIGskipamount}

\def\nopagenumber{\global\nopagenumbertrue}
\newif\ifnopagenumber \nopagenumberfalse
\footline={}
\headline={\ifnopagenumber {} \global\nopagenumberfalse
     \else{\hss\elevenrm-- \folio\ --\hss}\fi}

\catcode`\@=11
\def\vfootnote#1{\insert\footins\bgroup\tenpoint\baselineskip12pt
     \interlinepenalty\interfootnotelinepenalty
      \splittopskip\ht\strutbox 
       \splitmaxdepth\dp\strutbox \floatingpenalty=20000
        \parfillskip=0pt plus1fil \parindent=25pt
         \leftskip=0pt \rightskip=0pt \spaceskip=0pt \xspaceskip=0pt
          \textindent{#1}\footstrut\futurelet\next\fo@t}
\catcode`\@=12
\newcount\notenumber

\def\note{\global\advance\notenumber by 1 \footnote{$^{\the\notenumber}$}}

\def\titlepage{\hrule height 0pt \nobreak\vskip 0pt plus0.8fil
     \global\nopagenumbertrue \baselineskip20pt}
\def\endtitlepage{\vskip 0pt plus 1.3fil \normalbaselines \eject}
\def\title#1{\vskip 0pt plus .1\vsize
     \center{\baselineskip 20pt \BIGskip \let\smfont=\secfont \titlefont\ #1}}
\def\author#1{\center{\baselineskip 20pt \MEDskip \sc\ #1}}
\def\affiliation#1{\center{\baselineskip 20pt \vskip0pt \elevenrm\ #1}}
\def\date#1{\center{\baselineskip 20pt \BIGskip \it\ #1}}
\def\and{\baselineskip 20pt \MEDskip \center{\sc\ and}}


\def\abstract{\BIGskip \begingroup \centerline{ABSTRACT} \leftskip=20pt
     \rightskip=20pt \baselineskip=14pt \parindent=15pt \nobreak\SMALLskip
      \parskip=0pt}


\def\section#1\par{\vskip0pt plus.05\vsize\penalty-100
     \BIGskip \vskip\parskip
      \msg{#1}\center{\baselineskip 15pt \let\smfont=\tenbf
       \secfont\ #1}\nobreak\MEDskip}
\def\subsection#1\par{\vskip0pt plus.03\vsize\penalty-60 \MEDskip\vskip\parskip
     \msg{#1}\center{\baselineskip 15pt \let\smfont=\tensl
      \subfont\ #1}\nobreak\MEDskip}
\def\subsubsection#1\par{\vskip0pt plus.02\vsize\penalty-40 \SMALLskip\vskip
     \parskip\msg{#1}\center{\baselineskip 15pt \let\smfont=\qtensl
      \subsubfont\ #1}\nobreak\SMALLskip}



\def\bysame{\hbox to 80pt{\leaders\hrule height 2.4pt depth -2pt\hfill .\ }}

\def\bysame{\hbox to 80pt{\leaders\hrule height 2.4pt depth -2pt\hfill .\ }}


\def\figure#1{\par\hangindent40pt \noindent{Figure #1: }\nobreak}

 \def\chaphead{}

\newcount\tablenumber \newbox\tabbox \newdimen\tabwid
\def\clrtablenumber{\global\tablenumber=0} \clrtablenumber
\def\lasttable{{\rm\chaphead\the\tablenumber}\SP}
\def\nexttable{\advance\tablenumber by1 {\rm\chaphead\the\tablenumber}%
   \advance\tablenumber by-1 \SP}
\def\PM{\ifmmode\pm\else${}\pm{}$\fi}
\def\trule{\noalign{\vskip4pt}\noalign{\hrule}\noalign{\vskip4pt}}

\let\tablefont=\tenpoint
\font\sc=cmcsc10
\newdimen\tindent \newskip\tbaseskip \tbaseskip=12pt

{\catcode`\?=\active \catcode`\"=\active \catcode`\^^M=\active%
\gdef\table#1[#2]{\global\advance\tablenumber by1\begingroup%
  \ifx#1\empty\else\xdef#1{\chaphead\the\tablenumber\noexpand\SP}\fi%
  \tablefont \baselineskip\tbaseskip%
  \def\hhh{\hskip0pt plus1000pt}\def\?{\char`\?}%
  \def\a{\rlap{$^a$}} \def\b{\rlap{$^b$}} \def\c{\rlap{$^c$}}%
  \def\d{\rlap{$^d$}} \def\e{\rlap{$^e$}} \def\f{\rlap{$^f$}}%
  \def\g{\rlap{$^g$}} \def\h{\rlap{$^h$}} \def\i{\rlap{$^i$}}%
  \def\j{\rlap{$^j$}} \def\k{\rlap{$^k$}} \def\l{\rlap{$^l$}}%
  \def\|##1{\rlap{$^{\rm ##1}$}}%
  \def\ms##1{\multispan ##1}%
\def\...{\null\nobreak\leaders\hbox to0.5em{\hss.\hss}\hskip1.5em plus1filll\
}%
  \catcode`\"=\active%
  \catcode`\^^I=4 \catcode`\^^M=\active \catcode`~=\active \catcode`?=\active%
  \def?{\enspace} \def~{\hfil} \def^^M{\crcr} \def"{\quad}\tabskip=0pt%
  \pageinsert\vbox to\vsize\bgroup\hrule height0pt\vfil%
  \centerline{TABLE \chaphead\the\tablenumber}%
  \def\tablename{#2}\if\tablename\empty\else\centerline{{\sc\tablename}}\fi%
  \setbox\tabbox=\vbox\bgroup\halign\bgroup}%
\gdef\tabcont{\begingroup%
  \tablefont \baselineskip\tbaseskip%
  \def\hhh{\hskip0pt plus1000pt}\def\?{\char`\?}%
  \def\a{\rlap{$^a$}} \def\b{\rlap{$^b$}} \def\c{\rlap{$^c$}}%
  \def\d{\rlap{$^d$}} \def\e{\rlap{$^e$}} \def\f{\rlap{$^f$}}%
  \def\g{\rlap{$^g$}} \def\h{\rlap{$^h$}} \def\i{\rlap{$^i$}}%
  \def\j{\rlap{$^j$}} \def\k{\rlap{$^k$}} \def\l{\rlap{$^l$}}%
  \def\|##1{\rlap{$^{\rm ##1}$}}%
  \def\ms##1{\multispan ##1}%
\def\...{\null\nobreak\leaders\hbox to0.5em{\hss.\hss}\hskip1.5em plus1filll\
}%
  \catcode`\"=\active%
  \catcode`\^^I=4 \catcode`\^^M=\active \catcode`~=\active \catcode`?=\active%
  \def?{\enspace} \def~{\hfil} \def^^M{\crcr} \def"{\quad}\tabskip=0pt%
  \pageinsert\vbox to\vsize\bgroup\hrule height0pt\vfil%
  \centerline{TABLE \chaphead\the\tablenumber{ \it (continued)}}%
  \setbox\tabbox=\vbox\bgroup\halign\bgroup}%
\gdef\tnotes{\trule\egroup\egroup\tabwid=\wd\tabbox%
  \hbox to\hsize{\hfill{\box\tabbox}\hfill}%
  \def\a{{$^a$}} \def\b{{$^b$}} \def\c{{$^c$}}%
  \def\d{{$^d$}} \def\e{{$^e$}} \def\f{{$^f$}}%
  \def\g{{$^g$}} \def\h{{$^h$}} \def\i{{$^i$}}%
  \def\j{{$^j$}} \def\k{{$^k$}} \def\l{{$^l$}}%
  \def\|##1{{$^{\rm ##1}$}}%
  \baselineskip\footbaseskip%
  \advance\tabwid by-\hsize\divide\tabwid by-2\advance\tabwid by20pt%
  \leftskip=\tabwid%
  \rightskip=\tabwid \parskip=1pt \parindent=0pt \catcode`\^^M=5%
  \def\par{\endgraf\leftskip=\tabwid\rightskip=\tabwid\hangindent=10pt}%
  \def\endtable{\par\vfil\egroup\endinsert\endgroup}\hangindent=10pt}}%
\def\endtable{\trule\egroup\egroup\hbox to\hsize{\hfill{\box\tabbox}\hfill}
  \vfil\egroup\endinsert\endgroup}

\elevenpoint
\SMALLskipamount=4pt plus1pt
\MEDskipamount=8pt plus2pt
\BIGskipamount=16pt plus4pt
\footbaseskip=16pt
\parindent=30pt
\hsize 15 truecm
\vsize 20.5 truecm
\hoffset 0.8 truecm
\voffset 1.0 truecm
\hrule height0pt
\nopagenumber

\def\lax    {\ifmmode{_<\atop^{\sim}}\else{${_<\atop^{\sim}}$}\fi}
\def\gax    {\ifmmode{_>\atop^{\sim}}\else{${_>\atop^{\sim}}$}\fi}

\def\asec{$^{\prime\prime}$}
\def\deg{$^{\circ}$}

\def\h2{$\rm H_2$}
\def\nh2{n_{H_2}}



\centerline {\chapfont References}

\bigskip

\oneandhalfspace

\noindent
Andr\'e, P., \& Montmerle, T.\ 1994, ApJ, 420, 837

\noindent
Bastien, P. 1982, A\&AS, 48, 153

\noindent
Beckwith, S.\ V.\ W., \& Sargent, A.\ I.\ 1991, ApJ, 381, 250

\noindent
Beckwith, S.\ V.\ W., \& Sargent, A.\ I.\ 1993a, in Protostars
and Planets III, edited by E.\ H.\ Levy

and J.\ L.\ Lunine (University
of Arizona Press, Tucson) p.\ 521

\noindent
Beckwith, S.\ V.\ W., \& Sargent, A.\ I.\ 1993b, ApJ, 402, 280

\noindent
Beckwith, S.\ V.\ W., Sargent, A.\ I., Scoville, N.\ Z.,
Masson, C.\ R., Zuckerman, B.\ \& Phillips,

T.\ G.\ 1986, ApJ, 309, 755

\noindent
Beckwith, S.\ V.\ W., Sargent, A.\ I., Chini, R.\ \&G\"usten, R.\
1990, AJ, 99, 924

\noindent
Cohen, M., \& Kuhi, L.\ V.\ 1979, ApJS, 41, 743

\noindent
Cohen, M., Emerson, J.\ P.\ \& Beichman, C.\ A.\ 1989, ApJ, 339, 455

\noindent
d'Antona, F., \& Mazzitelli, I.\ 1993, ApJS, 90, 467

\noindent
de Geus, E.\ J., Bronfman, L., \& Thaddeus, P.\ 1990, A\&A, 231, 137

\noindent
Dutrey A., Guilloteau, S.\ \& Simon, M.\ 1994, A\&A, 286, 149

\noindent
Elias, J.\ 1978, ApJ, 224, 857

\noindent
Elsasser, H., \& Staude, H.\ J.\ 1978, A\& A, 70, L3

\noindent
Galli, D., \& Shu, F.\ H.\ 1993, ApJ, 417, 220


\noindent
Ghez, A., Neugebauer, G.\ \&  Mathews, K.\ 1993, AJ, 106, 2005

\noindent
Hartmann, L., Hewett, R., Stahler, S., \& Mathieu, R.\ D.\ 1986, ApJ,
309,275

\noindent
Hayashi, M., Ohashi, N.\ \& Miyama, S.\ M.\ 1993, ApJ, 418, L71

\noindent
Henning, T., Pfau, W., Zinnecker, H., \&  Prusti, T.\ 1994, A\&A, 276, 129

\noindent
Herbig, G.\ H., \& Bell, K.\ R.\ 1988, in Lick Observatory
Bulletin No. 1111 (University of California)

\noindent
Herbig, G.\ H.\ 1977, ApJ, 214, 747

\noindent
Hirth, G.\ A., Mundt, R., Solf, J., \& Ray, T.\ P.\ 1994, ApJ, 427, L99

\noindent
Leinert, Ch., Hass, M., Richi, A., Zinnecker, H., \& Mundt, R.\ 1991, A\&A,
250, 407

\noindent
Kawabe, R., Ishiguro, M., Omodaka, T., Kitamura, Y., \& Miyama, S.\ M.\ 1993,
ApJ, L63

\noindent
Kenyon. S.\ J., Calvet, N., \& Hartmann, L.\ 1993, ApJ, 414, 676

\noindent
Koerner, D.\ W., Sargent, A.\ I., \& Beckwith, S.\ V.\ W.\ 1993a, Icarus.
 106, 2

\noindent
Koerner, D.\ W., Sargent, A.\ I., \& Beckwith, S.\ V.\ W.\ 1993b, ApJ, 408, L93

\noindent
Nakajima, T., \& Golimowski, D.\ A.\ 1995, AJ, in press

\noindent
Sargent, A.\ I., \& Beckwith, S.\ 1987, ApJ, 323, 294

\noindent
Sargent, A.\ I., \& Beckwith, S.\ V.\ W.\ 1989, in
Structure and Dynamics of the Interstellar Medium,

IAU Colloquium No.\ 120, 215

\noindent
Sargent, A.\ I., \& Beckwith, S.\ 1991, ApJ, 382, L31

\noindent
Sargent, A.\ I., \& Beckwith, S.\ 1993, in Millimeter and Submillimeter
Wave Interferometry, edited

by M.\ Ishiguro and R.\ Kawabe,
(Bookcrafters, San Francisco) p.\ 232

\noindent
Scoville, N.\ Z., Sargent, A.\ I., Sanders, D.\ B., Claussen, M.\ J.,
Masson, C.\ R., Lo, K., \& Phillips,

T.\ G.\ 1986, ApJ, 303, 416

\noindent
Stahler, S.\ W., Korycansky, D.\ G., Brothers, M.\ J., \&
Touma, J.\ 1994, ApJ, 431, 341

\noindent
Stapelfeldt, K.\ R., Burrows, C.\ J., Krist, J., Hester, J.\ J.,
Trauger, J., Ballester, G., Casertano, S.,

Clarke, J., Crisp, D.,
Evans, R.\ W., Gallagher, J.\ S., Griffiths, R.\ E., Hoessel, J.\ G.,

Holtzman, J.\ A., Mould, J.\ R., Scowen, P., Watson, A.\ M., \&
Westphal, J.\ A.\ 1994,

ApJ, submitted

\noindent
Strom, K.\ M., Strom, S.\ E., Edwards, S., Cabrit, S., \&
Skrutskie, M.\ F. 1989, AJ, 97, 1451

\noindent
Tamura, M., \& Sato, S.\ 1989, AJ, 98, 1368

\noindent
Ungerechts, H., \& Thaddeus, P.\ 1987, ApJS, 377, 510

\noindent
van Langevelde, H.\ J., van Dishoeck, E.\ F., \& Blake, G.\ A.\
1994, ApJ,in Press

\noindent
Weintraub, D.\ A., Masson, C.\ R., \& Zuckerman, B.\ 1987, ApJ, 320, 336

\noindent
Weintraub, D.\ A., Masson, C.\ R., \& Zuckerman, B.\ 1989, ApJ, 344, 915

\noindent
Whitney, B.\ A., \& Hartmann, L.\ 1992, ApJ, 395, 529

\noindent
Whitney, B.\ A., \& Hartmann, L.\ 1993, ApJ, 402, 605

\noindent
Whittet, D.\ C.\ B.\ 1974, MNRAS, 168, 371


\font\qf=cmtt10 at 12truept

\titlepage
\title{Imaging the Small-Scale Circumstellar Gas Around T~Tauri
Stars}

\author{D. W. Koerner$^{1}$, \and A. I. Sargent$^2$}
\vskip 1.25truein

\affiliation{$^1$ Jet Propulsion Laboratory, 169-506,
4800 Oak Grove Dr., Pasadena, CA 91109.
Telephone: (818)354-7893; FAX: (818)354-8895;
E-mail: davidk@coma.jpl.nasa.gov}
\bigskip
\affiliation{$^2$ Division of Physics, Mathematics and Astronomy,
105-24, California Institute of Technology,
Pasadena, CA 91125. Telephone: (818)356-6622; FAX: (818)568-9352;
E-mail: afs@astro.caltech.edu}

\vskip 2.4truein
\noindent
No. pages: 32

\noindent
No. Figures: 5


\endtitlepage








\doublespace

\centerline{\bf Abstract}
\bigskip

We have detected circumstellar molecular gas around
a small sample of T Tauri stars through aperture synthesis imaging of
CO(2$\rightarrow$1) emission at $\sim$~2-3$''$ resolution.
RY Tauri, DL Tauri, DO Tauri, and AS 209 show resolved
and elongated gaseous emission.  For RY Tau, the deconvolved,
half-maximum radius along the direction of elongation, PA$\sim$48$^\circ$,
is 110 AU. Corresponding radii and orientations for the other
sources are: DL Tau -- 250 AU at PA$\sim$84$^\circ$;
DO Tau -- 350 AU at PA$\sim$160$^\circ$;
AS 209 -- 290 AU at PA$\sim$138$^\circ$.
RY Tau, DL Tau, and AS 209 show velocity gradients parallel to
the elongation, suggesting that the circumstellar
material is rotating.
RY Tau and AS 209 also exhibit double-peaked spectra
characteristic of a rotating disk. Line emission
from DO Tau is dominated by high-velocity blue-shifted gas
which complicates the interpretation.
Nevertheless, there is in each case sufficient evidence to
speculate that the circumstellar emission may arise from
a protoplanetary disk similar to that from which our solar
system formed.

\newpage

\centerline{\bf 1. Introduction}
\medskip

There is increasing evidence that the nebular environment
assumed to have given rise
to Solar System bodies is a common feature of
the early stages
of stellar evolution. Circumstellar dust disks are
detected around a large
fraction of T Tauri stars (e.g., Cohen, Emerson, \& Beichman 1989;
Strom {\it et al.\ }1989; Beckwith
{\it et al.\ }1990; Andr\'e \& Montmerle 1994;
Henning {\it et al.\ }1994) and in at least one case, the
velocity field of the associated circumstellar gas
is consistent with that of a model
of a disk in Keplerian rotation (Koerner, Sargent,
\& Beckwith 1993a, henceforth KSB).  Disk properties, such
as mass, size, and kinematics, are sufficiently similar to
those expected for the early solar nebula (cf.\ Beckwith \&
Sargent 1993a) that continued study is likely to
afford us the opportunity to constrain theories of
planetary system formation.  A better understanding
of the detailed properties of these circumstellar
environments may also narrow down the field of
truly proto-planetary disks.

The small fraction of T Tauri stars for which circumstellar gas has
been observed (cf.\ Sargent \& Beckwith 1993)
contrasts markedly with the 50\% detection rate
achieved by continuum surveys. This is largely because
single telescope observations of the circumstellar dust continuum
emission are readily obtainable. In general,
unambiguous detection of circumstellar gas
on small spatial scales requires molecular line millimeter-wave
interferometer measurements; historically, these have been
time consuming. Flattened structures of circumstellar gas with radii of order
1000 AU have been observed around the T Tauri stars,
HL Tauri (Beckwith {\it et al.\ }1986; Sargent \& Beckwith 1987),
T Tauri (Weintraub {\it et al.\ }1987; 1989),
DG Tauri (Sargent \& Beckwith 1989; Ohashi {\it et al.\ }1991),
GG Tauri (Kawabe {\it et al.\ }1993; Koerner, Sargent, \& Beckwith
1993b; Dutrey, Simon, \& Guilloteau 1994) and GM Aurigae (KSB).
The earliest interpretations suggested that circumstellar material
was centrifugally supported out to radial distances of 1000 AU
(Sargent \& Beckwith 1987; Weintraub {\it et al.\ }1989).
Recent observations indicate that gas at these large radii is still
in a state of collapse (e.g., Hayashi {\it et al.\ }1993; van Langevelde,
van Dishoeck, \& Blake 1994). However, it also appears that some of
these objects may not represent most T Tauri stars in general.
For example,
Hubble Space Telescope observations
imply that HL Tau is heavily embedded in its parent
cloud and is therefore at a very early stage of pre-main
sequence evolution (Stapelfeldt {\it et al.\ }1994);
the spectral energy distribution (SED) of T Tau, a binary object,
can be understood only if there is a large
component of non-planar dust
(Kenyon {\it et al.\ }1993),
and GG Tau is part of an unusual quadruple star system and surrounded
by a large torus (Leinert {\it et al.\ }1991;
Ghez {\it et al.\ }1993; Koerner
{\it et al.\ }1993b; Dutrey {\it et al.\ }1994).

To date, GM Aur seems the best example of a ``typical''
T Tauri star. It is relatively
unobscured with A$_V$ = 0.1
(Strom {\it et al.\ }1989) and is significantly
older than the other stars which show circumstellar gas.
It is also not a binary and affords a closer example
of the early Solar System environment than the large
fraction of T Tauri stars with a stellar companion
(Ghez {\it et al.\ }1993).
Here, there is unambiguous evidence that the
surrounding molecular gas is centrifugally supported out
to 1000 AU, but emission beyond
a 150 AU radius is tenous compared to
the other circumstellar structures observed.
This star/disk system may well represent a planet-forming
environment (KSB).  Obviously, we would like to observe a range
of circumstellar gaseous environments for a larger sample
of T Tauri stars to determine detailed properties of the disks and generalize
these to compare with the conditions in the early solar nebula.
We are continuing a program of aperture synthesis imaging of the environments
of T Tauri stars with a view to building up a statistical base of
information on the circumstellar gas. High resolution molecular line
images of four more systems, RY Tauri, DL tauri, DO Tauri, and AS 209 are
presented and discussed here.

\medskip
\centerline{\bf 2. Observations}
\medskip

RY Tau, DL Tau, DO Tau, and AS 209
are among the strongest sources of
millimeter continuum emission in the nearest star-forming regions
(Beckwith {\it et al.\ }1990; Andr\'e \& Montmerle 1994).
Their coordinates  (Herbig \& Bell 1988) and relevant
properties are listed in Table 1. The spectral index,
$\alpha_{2.2-12\mu}$, was derived from
K-band magnitudes (Cohen \& Kuhi 1979; Ghez 1993)
and IRAS 12 $\mu$m fluxes (Weaver \& Jones 1992).
The Taurus sources are located at distances of 140 pc and
AS 209, near the Ophiuchi dark cloud, is at 160 pc
(Elias 1978; Whittet 1974; de Geus
{\it et al.\ }1990).

Observations of  RY Tau, DO Tau, DL Tau, and AS 209
in the 1.3 mm continuum and CO ($\2to1$) line
at 230.5 GHz were made between 1993, November, and 1994, March, using
the five-element Owens Valley millimeter-wave array.  Baselines up to 60 m
E--W and 60 m N--S provided an approximately circular
$3''$ beam (FWHM) with area $\Omega_B$ =
6.9 $\times$ $10^{-10}$ sr for DO Tau and
$\Omega_B$ = 6.1 $\times$ $10^{-10}$ sr for
AS 209. More extended baselines, up
to 100 m E--W and N--S, yielded spatial resolution of up to 2$''$
for RY Tau and DL Tau with $\Omega_B$ =
3.6 $\times$ $10^{-10}$ sr and $\Omega_B$ = 6.0 $\times$ $10^{-10}$ sr,
respectively. Cryogenically-cooled SIS receivers on each
telescope produced over-all system temperatures of
500--1000 K. The digital correlator was
configured to have two bands of Hanning smoothed channels,
32 $\times$ 1 MHz and 96 $\times$ 83 kHz,
yielding spectral resolutions 1.30 and
0.11 km s$^{-1}$, respectively. Each band was
centered at $V_{LSR} \sim$ 6.0--7.5 \kms. for the
various Taurus sources (cf.\ Hartmann {\it et al.\ }1986)
and $V_{HEL} = -4.0$ \kms. for AS 209.
Continuum measurements were made
simultaneously in a broad-band channel of width 1 GHz.
The quasar, 0528+134, was observed at 20-minute intervals
to calibrate visibility phases for DO, DL, and RY Tau;
NRAO 530 was used for AS 209. The absolute flux
density scale relied on measurements of Uranus and
3C 273. Maps made with the NRAO AIPS software package
are centered on the positions in Table 1.
Uncertainties in absolute fluxes and
positions are estimated to be 15$\%$ and 0$''$.5, respectively, for
peak map intensities, but may as high as 50$\%$ and one beamwidth
($\sim$ 2--3$''$) for features detected at 3--5 $\sigma$ levels.

\bigskip
\centerline {\bf 3. Results}

Maps of
the integrated CO($\2to1$) emission for each source
are displayed in Fig.\ 1a--4a. Values for the integrated
intensity are given in column (2) of Table 2.
For RY Tau and DL Tau,
integration was over the velocity range for which emission appeared
above a level of 300 mJy beam$^{-1}$ (2$\sigma$) at spectral
resolution 0.45 \kms..  Since
the linewidths for AS 209 and DO Tau exceeded the velocity
range of the high-velocity-resolution band, emission was integrated
over the wider 1.30 \kms. channels
in which peak map intensity was above a level of
50 mJy beam$^{-1}$ (2$\sigma$).
Individual velocity ranges
are listed in column (3) of Table 2.

\def\trans2{${\rm CO}$(2$\rightarrow$1)}

\medskip
\noindent
{\bf 3.1 Morphology, Optical Depth, and Mass}

To obtain the first approximate determination of the
radial extent of the circumstellar gas,
elliptical Gaussians were fit to the images shown in Fig. 1a--4a
and deconvolved with the
restoring beams from the CLEAN deconvolution
process. The radius at half-maximum intensity, R$_{HM}$, in
column (4) of Table 2 is the FWHM major axis of this Gaussian fit.
If the expected power-law profiles in temperature and density
prevail, emission from the outer parts of disks is drastically reduced
compared to the central regions. Thus,
$R_{HM}$ is a {\it lower limit} to the full radial extent of the gas.
All of the sources appear to be
resolved in  $\sim$2-3$''$ beams, with
100 AU $\simless$ R$_{HM}$ $\simless$ 400 AU.
These dimensions are closer to those anticipated for the early
solar nebula and approximately
an order of magnitude lower than
measured for other T Tauri stars.

As shown in KSB, an upper limit for the size of
a central optically thick region in a circumstellar disk
can be derived from the brightness temperature of peak
emission at the central velocity.
Assuming a typical disk temperature of $T_k$ = 30 K
(Beckwith \& Sargent 1993b), an
inclination angle of 45$^\circ$, and taking into account the
average beam size ($\sim 3''$), a limiting radius
for an optically thick region for sources at distances of 150 pc
is given by:
$$ R_{\tau>1} = 80 [B]^{1/2} AU $$
where B is the line-center intensity of peak
emission (in Jy beam$^{-1}$)
in a 1.30 \kms. wide velocity bin. This is a conservative
upper limit, since contributions from optically thin emission are ignored,
and higher temperatures will significantly lower the result.
For objects where a sizeable fraction of circumstellar
mass is still infalling, the average temperature is
likely to be higher.

In column (5) of Table 2,
R$_{\tau>1}$ is listed for each source, taking into account
individual beam sizes and subtracting any contribution from
the peak continuum flux (never more than $20$\%).
A comparison with R$_{HM}$  reflects the
extent to which the emission is optically thin.
As will be evident, this is important in estimating
the circumstellar mass from the
CO($\2to1$) fluxes.

The total mass of circumstellar gas is calculated from the
integrated CO($\2to1$) emission, ${\int S_\nu dv}$,
using (cf. Scoville et al.\ 1976):

$$M_{H_{2}} = 1.42~\times~10^{-10}
	~{(T_x + 0.93) \over e^{-16.76/T_x}}
	~{\tau_{CO} \over (1-e^{-\tau})} ~{D_{kpc}^2 \over
	 X(CO)} {\int S_\nu dv} \quad \quad M_\odot.$$
\medskip

\noindent
Here, $T_x$ is the excitation temperature, $\tau_{CO}$
is the optical depth in the CO line, $D_{kpc}$ is the distance to the
source in kpc, and $X(CO)$ is the fractional abundance of CO.
For optically thin emission,

$$~{\tau_{CO} \over (1-e^{-\tau})}~\approx~1\ ,$$

\medskip
\noindent
and the mass is calculable without {\it a priori} knowledge of $\tau_{CO}$.
If the magnitude of the optical depth is completely unknown,
the above approximation allows a determination of
a lower limit to $M_{H_2}$.

Masses, $M_{H_2}$, were calculated for all the sources
using the integrated
line fluxes listed in Table 2, adopting an average value of
$T_x = 30$K,
$X(CO) = 10^{-4}$, $D_{kpc} = 0.140$ kpc for the Taurus sources,
and $0.160$ kpc for AS 209.
The results are shown in column (6) of Table 2.
As optical depth decreases,  $M_{H_2}$
approaches the true mass;
$R_{\tau>1}$/R$_{HM}$ provides a qualitative assessment
of the accuracy of the estimate.
Table 2 shows that, for the entire sample, the upper limit to
the radius of an optically thick region equals or exceeds
the size of the present solar system (50 AU).
Total masses much larger than the limits in Table 2, and still
confined within a solar-system-sized radius,
could easily be consistent with
these observations. Indeed, masses typically 2 orders
of magnitude larger are implied by
1.3 mm continuum measurements (Beckwith {\it et al.\ }1990;
Andr\'e \& Montmerle 1994).  Further observations in
optically thin isostopes of CO are needed to
refine estimates of the mass of circumstellar gas.

\medskip
\noindent
{\bf 3.2 Orientation and Kinematic Implications}

The true orientation of the elongation axis
of the circumstellar gas is critical to any
kinematic analysis. In Fig. 1a--4a,
emission is preferentially elongated along one axis,
implying a flattened structure, but
the degree of elongation
is uncertain, since some objects are very small compared to the
spatial resolution of the maps.  However, the orientation of
an associated optical jet can also be used
to infer the PA of the sky-projected axis normal to the disk plane.

In Table 3, the PA of CO emission elongation, $\Theta_{CO}$,
determined from the Gaussian fit, is listed for each of the sources.
Uncertainties tabulated for $\Theta_{CO}$ are derived from the
goodness of fit and do not reflect the amplitude and position
uncertainties in the map itself.  Errors for RY Tau, where the
deconvolved size is less than the FWHM size of the synthesized beam, are
likely to be greater than the formal uncertainties listed
in Table 3. In addition, irregular source features can
confuse the determination
of elongation orientation, especially when spatial resolution
is limited. These features include
azimuthal asymmetries in the circumstellar gas itself,
molecular outflows, or ambient cloud material.
Independent ways of constraining the orientation of the
circumstellar material must be employed. In fact,
the spatial distribution of the
intensity-weighted average velocity  (``first
moment'' with respect to velocity)
can provide circumstantial evidence about disk orientation, if
systematic velocity gradients are apparent.
Maps of the first moment in velocity for the sample sources
are shown in Fig.\ 1b--4b.

To a first approximation, rotation in a flattened
disk will appear as a gradient in mean velocity
parallel to the elongation axis of the disk.
Outflow, or pure infall in the plane of a
flattened envelope, on the other hand, will result
in a gradient orthogonal to this direction.
Complex mean velocity structure will result
from a combination of these effects, and
the velocity gradient alone is not an
unambiguous indicator of disk orientation,
especially if the elongation itself is due to outflow.
However, the molecular gas in maps shown here
exhibits much smaller linewidths than
typically observed for outflow sources (except for DO Tau)
and is positioned close to the star.
Therefore, for purposes of preliminary interpretation,
the coincidence of a mean velocity gradient with
the apparent CO elongation axis is considered as
a significant clue to disk orientation.
More detailed kinematic modeling
like that carried out for GM Aur by KSB will be
presented in a future paper.
The PA of any systematic velocity gradient seen in
Fig.\ 1b--4b, $\Theta_{{\delta}V}$,
is given in column (3) of Table 3
for comparison with $\Theta_{CO}$.
For RY Tau and AS 209, $\Theta_{{\delta}V}$
was measured by taking the perpendicular to contours at the position
of peak emission in Fig. 1a and 2a, since
curvature was evident in the velocity gradient.

Other constraints on the orientation
of a possible disk include the observation of
an optical jet, presumably oriented perpendicular to
the axis of elongation. The implied disk orientation
is given as $\Theta_{jet}$ in Table 3.
Although $\Theta_{jet}$  cannot be used alone to define
orientation, agreement between several indicators
strongly suggests the correct orientation of planar
material.  The PA of spatially
averaged polarization vectors of scattered
stellar light, $\Theta_{pol}$, is also a function of the orientation of
the scattering material (Els\"asser \& Staude 1978;
Bastien 1982).
Monte Carlo models of light scattered
by disks and envelopes indicate that $\Theta_{pol}$
will be parallel to the elongation axis for flared disks
and envelopes, but perpendicular to it for thin
disks (Whitney \& Hartmann 1992; 1993).
In Table 3, the agreement between $\Theta_{CO}$,
$\Theta_{\delta{v}}$, and $\Theta_{pol}$
is generally quite good and is supported by $\Theta_{jet}$
in the case of DO Tau (Hirth {\it et al.\ }1994).
We claim this greatly strengthens an interpretation of the
structures in Fig. 1--4 as thick disks or flattened envelopes.
Thin disks are ruled out, since $\Theta_{pol}$ would be
expected to differ from $\Theta_{CO}$ and $\Theta_{\delta{v}}$
by $90^\circ$.
\bigskip
\noindent
{\bf 3.3 Velocity Structure of Circumstellar Gas}
\medskip

Maps of the first moment with respect to velocity
for RY Tau, DL Tau, DO Tau, and AS 209,
in Fig.\ 1b--4b, display
velocity gradients parallel to the elongation axis of CO emission
and strongly suggest the existence of rotating disks.
This evidence is not sufficient
to uniquely establish the kinematic structure of the gas, however,
since detailed information about the velocity distribution
of emission has been lost by averaging. In Fig.\ 5,
spatially integrated CO($\2to1$) spectra provide
a complementary presentation of the data which retains
velocity information. The emission intensity in spectral
line maps was integrated over a $5''$-radius region
centered on the stellar position and plotted as a function
of velocity to effectively separate the signature of circumstellar
gas from that of the large ambient cloud.

The spectra in Fig.\ 5 may be used to refine
preliminary interpretations of the circumstellar CO emission.
Two of the sources, RY Tau and AS 209, have lineshapes
which are approximately double-peaked,
consistent with Keplerian rotation (Beckwith \& Sargent 1993b).
This interpretation is somewhat uncertain for
RY Tau, since the peak-to-trough  amplitude is only
a few times that of the greatest channel-to-channel variations due
to noise. However, it is bolstered by available observations of
the stellar radial velocity; the molecular spectrum is centered
exactly on the optical value given by Hartmann et al.\ (1986),
$V_{LSR}$ = 7.8$\pm2.4$ \kms..  The line shape for AS 209
is double-peaked about a central velocity $V_{HEL}\approx-7$ \kms.,
consistent with a stellar velocity of
$V_{HEL}=-6\pm3$ \kms. (Walter, F.\ M., private communication).
The spectral line for DL Tau is centered at $V_{LSR} \approx$ 7 \kms., a
typical velocity for T Tauri stars in the
Taurus-Auriga molecular dark cloud (cf.\ Hartmann et al.\ 1986).
The linewidth is surprisingly narrow ($\Delta$V = 1.2 \kms.)
such that any
kinematic signature is likely to be masked by thermal broadening.

The remaining source, DO Tau, shows an extremely asymmetric
spectral line that is an order of magnitude stronger than the others.
Most of the molecular
emission is blue-shifted more than 10 \kms. relative to the velocity of
the ambient molecular cloud and the best available stellar velocity
($V_{LSR}=10\pm5$ \kms.; Herbig 1977). The dominant blue-shifted wing
is most likely due to a molecular outflow,
especially since a blue-shifted optical jet is known
to emanate from the central star (Hirth {\it et al.\ }1994).
Redward of the peak emission, however,
there is a narrow feature separated from the blue-shifted emission
by a prominent dip at the systemic velocity of the local
molecular cloud.  These features may be
circumstellar in origin.

A ``double-peaked''
line shape arises from a centrifugally supported
disk as a result of the $1/\sqrt{r}$ dependence
of the Keplerian velocity, the $r^2$ dependence of emitting areas
with the same velocity, and the truncation of the disk at
some outer radius.  Peaks in the spectra largely reflect the
Keplerian velocities at the outer radius of the disk, modulated
by sin($i$), where $i$ is the angle of inclination to the line of
sight.  However, other spatial and kinematic configurations can
yield a similar lineshape. Line-center absorption by a cold intervening
molecular cloud can introduce a central dip in the lineshape,
especially in $^{12}$CO observations where the emission is likely to
be optically thick, and infall in a {\it flattened}
structure also produces a double-peaked line
shape, since infall velocities are also
proportional to $1/\sqrt{r}$.

Although none of the spectra in Fig.\ 5
can be taken as unambiguous proof
of the presence of a Keplerian disk,
they do provide a consistency check. The Keplerian velocity at the outer
radius, $\sqrt{GM/R_d}$, can be calculated assuming the central
mass is approximately the stellar mass,
$M \approx M_{_{\bf *}}$, listed in Table 1,
and that the outer disk radius is given by
$R_d \approx$ R$_{HM}$ of Table 2.
The derived Keplerian velocity can then be modulated over
a range of acceptable inclination angles and compared with
double-peak velocities suggested by
Fig.\ 5.  If the center-to-peak velocity
difference is significantly greater than that predicted for an
edge-on Keplerian disk of the observed size,
Keplerian rotation can be ruled out as the dominant
cause of the spectral lineshape.

Table 4 lists Keplerian velocities calculated in this
way and center-to-peak velocity differences estimated from the spectra in
Fig.\ 5. For RY Tau and AS 209, this was simply taken to
be half the distance between the channels with the highest
amplitudes.  For DL Tau, where the narrowness of the line obscures
any double-peaked structure, a nominal
value of one quarter of the total line width was chosen.
In the case of DO Tau, the distance between the
spectral dip and the narrow velocity feature was used.
None of the velocities estimated from the spectra exceed those of the
Keplerian approximation. In fact, since
free fall velocities exceed the Keplerian velocity by
a factor of $\sqrt{2}$,
this is more consistent with a rotational rather than an
infall interpretation.
The range of inclination angles required for agreement with
Keplerian rotation is given in Column 4. The values are plausible
and, in the case of the best-resolved disk, consistent with an
estimate based on the aspect ratio of the major and minor
axes of CO emission.

\medskip
\noindent
{\bf 3.4 Combined Evidence for Rotating Disks around Individual Stars}

None of the independent diagnostics presented thus far
demonstrate conclusively that the objects observed
are protoplanetary disks.  However, their combined effect
provides useful criteria with which to test the simplest possible
interpretations. These include a rotating disk, infall envelope,
bipolar outflow, or some combination of these;
more complicated explanations can perhaps be constructed, but
their consideration is unwarranted unless these hypotheses are
ruled out. With these restrictions in mind,
it is useful to briefly summarize the diagnostics for each individual star
in order to assess the likelihood that observed CO emission
arises from a protoplanetary disk.
\medskip

\noindent
{\it RY Tau:} Circumstellar CO emission around RY Tau is compact,
located at the stellar position, and exhibits
a symmetric double-peaked
lineshape centered at the radial stellar velocity.
The linewidth is consistent with that from a
disk inclined 25$^\circ$ from face on
with radius $\sim$100 AU and in Keplerian rotation
about a central mass dominated by that of the star.
Indicators of orientation  --  the PA's of CO elongation, velocity gradient,
and polarization -- also support a rotating-disk interpretation
and are aligned within 20$^\circ$ of each other.
The observed values are somewhat uncertain due to the small
radial extent of CO and variability of polarization, but
their validity is confirmed by the orientation
of a reflection nebulosity associated with RY Tau
in R-- and I--band coronagraphic images
(Nakajima \& Golimowski 1995).
Furthermore, the compact size, lineshape, and linewidth
argue that the CO is truly circumstellar and
not part of an outflow.  It must be conceded,
however, that an element of infall
is still possible, and a compact low-velocity outflow
is not strictly excluded.
\medskip

\noindent
{\it DL Tau:} Orientation indicators for circumstellar
gas around DL Tau are in exact agreement and strongly
support the idea that CO emission
represents a 250-AU-radius circumstellar disk.
The spectral line is
centered at a typical velocity for T Tauri stars
in the Taurus-Auriga molecular dark cloud
and is surprisingly narrow. A nearly face-on disk
orientation is required to match it, given the
stellar mass estimate of 0.56 \msun..  However,
pre-main sequence stellar mass estimates are most
uncertain for stars of very low mass (cf.\ d'Antona \&
Mazzitelli 1993), and DL Tau is the least massive of the
sample. It
is also possible that the observed linewidth has been
partially narrowed by absorption from the ambient molecular cloud.
In either case, the narrow linewidth and compact morphology argue
against an outflow interpretation and the orientation diagnostics
overwhelmingly support a disk-like configuration. If this is truly
the case, infall is strictly excluded by the orientation
of the velocity gradient with respect to that of the elongation axis.
In these respects, the rotating disk interpretation is robust
for DL Tau. To confirm this, imaging at higher spatial resolution
is needed to resolve velocity-dependent
structure for comparison to a kinematic model.
\medskip

\noindent
{\it DO Tau: }
The orientation of circumstellar gas around DO Tau is well established
by the morphology of CO emission and observations of its optical jet
and polarization.
However, a single systematic velocity gradient is not apparent in the
first moment velocity map. The lineshape
is much broader than expected for rotation and asymmetrically
blue-shifted to velocities higher than would be expected for
Keplerian rotation in a disk that size.
Its association with an asymmetric optical jet virtually
establishes an outflow origin for
the high-velocity molecular emission.
In contrast, a disk-like appearance
and agreement between three orientation indicators argue for a
large circumstellar component of rotating and/or infalling gas.
Spectral features near the systemic
velocity are consistent with Keplerian rotation, but detailed study of the
emission morphology at these velocities is required to
identify any rotating disk. Clearly, some combination of
hypotheses is needed to explain the morphology and kinematics of
DO Tau's CO emission. To obtain the correct interpretation,
modeling of emission in individual spectral line
maps is being carried out and will be presented
in a future paper.
\medskip

\noindent
{\it AS 209:}
The map of CO emission reveals an elliptical shape
with somewhat tenuous extensions perpendicular to the elongation
axis. The dominant velocity gradient and polarization vector are parallel
to this axis, but the velocity gradient is not as clear as for
DL Tau and RY Tau. The lineshape
is double-peaked about a central velocity that agrees
approximately with estimates of the stellar velocity, and the
linewidth is consistent with a Keplerian disk inclined
$\sim25^\circ$ from edge on. Taken together, these
facts suggest that the emission is dominated by a rotating disk.
However, the perpendicular extensions, irregular velocity gradient,
and a slight asymmetry in the lineshape could easily be accounted for
by an additional outflow component to the molecular emission.
As for DO Tau, modeling of high spatial- and velocity-resolution
spectral line maps is required to confirm this picture and will be
the subject of a future paper.

\bigskip
\centerline
{\bf 4. Summary and Discussion}
\medskip

Four T~Tauri stars, RY Tau, DL Tau, DO Tau, and AS 209, have been
imaged in the CO$(\2to1$) line at $\sim$2--3$''$ resolution.
In all cases, emission is
resolved with deconvolved half-maximum radii of 100 to 350 AU.
The PA's of optical linear polarization
and, in one case, an optical jet show
remarkable agreement with disk orientation implied by the
CO elongation and by velocity gradients.
{}From the peak strength
of line emission, upper limits on the size of any optically thick inner region
are found to be 50--150 AU, the approximate size of the solar system.
Masses of circumstellar
H$_2$ range over 2 orders of magnitude, from 1.3 $\times 10^{-6}$
to 1.4 $\times 10^{-4}$ \msun., but these values are lower limits and
much less than
masses estimated from the dust
continuum emission (Beckwith {\it et al.\ }1990;
Andr\'e \& Montmerle 1994). Except for DO Tau, all sources
show evidence of a velocity gradient along the elongation
axis and spectral lineshapes and linewidths which
are consistent with Keplerian rotation about the estimated stellar
mass.

The strongest case for a centrifugally supported
disk can be made for circumstellar gas around DL Tau.
With RY Tau, it shows the most distinct
velocity gradient in first moment velocity maps.
For both AS 209 and RY Tau, however,
the uncertainty in orientation admits a possible
infall or outflow component to circumstellar gas. Nevertheless,
the PA of elongation is not likely to be off by as much as
45$^\circ$, indicating that the {\it dominant}
velocity component of the molecular gas is rotational.
Spectral lineshapes for all three objects are also
consistent with that of a Keplerian disk and strengthen
the case for a rotational interpretation of
the velocity gradients observed in
moment maps.  If the possible kinematic
interpretations are limited to rotation, outflow, or infall,
evidence presented here weighs heavily in favor
of predominantly rotation of circumstellar gas for
all three of these sources.

For DO Tau, the first moment velocity map and associated spectrum
eliminates the possibility that {\it all} of the observed emission
radiates from rotating circumstellar gas. An outflow interpretation
for some of the gas is supported by the presence of
an optical jet (Hirth {\it et al.\ }1994).
It is somewhat surprising,
therefore, that the morphology of molecular emission in Fig.\ 3
resembles a disk with orientation perpendicular to the jet.
The disk-like morphology and complex velocity
structure may indicate the presence
of a radially infalling component to the
circumstellar velocity structure
(e.g.\ Hayashi {\it et al.\ }1993; for theoretical
justification of a {\it flattened} infall region, see Galli \& Shu 1993
and Stahler {\it et al.\ }1994). Molecular material around DO Tau,
with R$_{HM}$ = 350 AU, is substantially
more extended than that around the other sources,
but is still more compact than
for many earlier objects such as HL Tau.  It may represent an
intermediate evolutionary state in which much of the
circumstellar material is centrifugally supported, but
some is still in a state of infall.
Modeling of high-spatial-resolution spectral line maps of DO Tau
with a kinematic model of infall and rotation is required to
identify such a kinematic state (Koerner 1994) and will be
presented in a future paper.

In the above respects, circumstellar CO($\2to1$) emission from
the T~Tauri stars surveyed shows characteristics which
confirm the hypothesis that these stars are surrounded by
protoplanetary disks like that which gave rise to our solar system.
Further work is needed to secure this interpretation.
Kinematic modeling of spectral line maps of
\13CO.($\2to1$) emission from GM Aurigae provided
satisfactory evidence for
Keplerian rotation of circumstellar gas by duplicating
systematic patterns of emission seen in maps of the
disk at discrete velocities (KSB).
However, the circumstellar gas structures described here
are much smaller, and similar analyses
will require imaging at correspondingly higher spatial
resolution.

Despite the inability to rigorously identify kinematic patterns,
this work confirms the presence of molecular gas in the circumstellar
environment of T Tauri stars for which dust continuum emission is
detected at millimeter wavelengths. The radial extent of the gas
is more in accord with the dimensions of the solar system
than that of previously imaged T~Tauri stars;
molecular line emission is resolved in 2--3$''$
FWHM beams (corresponding to 300--450 AU) and optically thick
beyond a radial distance corresponding to the semi-major
axis of Pluto's orbit ($\sim$50 AU).
Some evidence of centrifugal support is present for most of the sample.
These facts strongly support the hypothesis that
a large fraction of T Tauri stars are surrounded by protoplanetary disks
similar to the early solar nebula.


\bigskip
\centerline
{\bf 5. Acknowledgments}  We thank the anonymous
referees for comments which helped clarify the results of this
paper. D. Koerner acknowledges support from NASA grant NGT-51071
and a Resident Research Associateship from the National Research
Council.
\medskip

\vfill
\eject

\vfill
\eject

\pageinsert
\bigskip
\bigskip
\vfil
\bigskip
\centerline {\bf Table I. Star Properties}
\bigskip
\hrule
\smallskip
\hrule
\medskip
\tabskip=1em plus2em minus0.5em
\halign to \hsize
{\noindent
      # & \hfil # & \hfil # & \hfil # & \hfil  # & \hfil # & \hfil # \cr
\noalign{\medskip}
\hfil (1) \hfil & (2) \hfil & (3) \hfil & (4) \hfil &  (5) \hfil
 & (6) \hfil & (7) \hfil \cr
\noalign{\smallskip}
\hfil Source \hfil & $\alpha$ \hfil & $\delta$ \hfil &
$F_{1.3mm}$  \hfil & $\alpha_{IR}$ \hfil & $M_{_{\bf *}}$ \hfil &
Age  \hfil \cr
\noalign{\smallskip}
        &  (1950) \hfil & (1950) \hfil & (mJy) \hfil & (2.2-12$\mu$m) \hfil
& (\msun.) \hfil & ($\times 10^6$ yr) \cr
\noalign{\medskip}
\noalign{\hrule}
\noalign{\bigskip}
RY Tauri & 04:18:50.80 & +28 19 35.0  & $229^a$ \hfil & $\ 0.05$ \hfil &
$1.69^a$
\hfil
& $0.21^a$ \cr
DL Tauri & 04:30:36.02 & +25 14 24.0  & $230^a$  \hfil & --0.37
\hfil & $0.56^a$ \hfil
& $1.20^a$ \cr
DO Tauri & 04:35:24.18 & +26 04 55.2  & $136^a$  \hfil & --0.33
\hfil & $0.72^a$ \hfil
& $0.60^a$ \cr
AS 209 & 16:46:25.24 & --14 16 56.5  & $300^b$  \hfil & --0.60
\hfil & --- \hfil & \hfil --- \hfil \cr
}
\bigskip
\hrule
\smallskip
\hrule
\bigskip
{\qf a) Beckwith {\it et al.\ }(1990)}
\medskip
{\qf b) Andr\'e \& Montmerle (1994)}
\bigskip
\vfil
\endinsert
\vfill
\eject

\pageinsert
\bigskip
\bigskip
\bigskip
\vfil
\centerline {\bf Table II. Global Properties Derived from CO Emission.}

\bigskip
\hrule
\smallskip
\hrule
\medskip
\tabskip=1em plus2em minus0.5em
\halign to \hsize
{\noindent
      # & \hfil # & \hfil # & \hfil # & \hfil  # & \hfil # \cr
\noalign{\medskip}
\hfil (1) \hfil & (2) \hfil & (3) \hfil & (4) \hfil &  (5) \hfil
 & (6) \hfil \cr
\noalign{\smallskip}
\hfil Source \hfil & $\int {S_\nu} dv$ \hfil & $\Delta$V \hfil &
R$_{HM}$ \hfil &
$R_{\tau>1}$ \hfil & $M_{H_2}$ \hfil \cr
\noalign{\smallskip}
        &  (Jy km/s) \hfil & (km/s) \hfil & (AU) \hfil & (AU) \hfil
& ($\times 10^{-6}$ \msun.) \hfil \cr
\noalign{\medskip}
\noalign{\hrule}
\noalign{\bigskip}
RY Tauri & 6.6 \hfil & 6.9 \hfil & 107 & 52  & 9.6 \hfil \cr
DL Tauri & 0.9 \hfil & 1.2 \hfil  & 250 & 53 & 1.3 \hfil \cr
DO Tauri & 100.2 \hfil & 13.6 \hfil & 350 & 164 & 145  \hfil \cr
AS 209 & 7.4 \hfil & 10.4 \hfil & 287 & 115 & 10.7 \hfil \cr
}
\bigskip
\hrule
\smallskip
\vfil
\endinsert

\vfill
\eject

\pageinsert
\bigskip
\bigskip
\bigskip
\vfil
\centerline  {\bf Table III. Diagnostics of Disk Orientation.}

\medskip
\hrule
\smallskip
\hrule
\medskip
\tabskip=1em plus2em minus0.5em
\halign to \hsize
{\noindent
      # & \hfil # & \hfil # & \hfil # & \hfil # \cr
\noalign{\medskip}
\hfil (1) \hfil & (2) \hfil & (3) \hfil & (4) \hfil & (5) \hfil \cr
\noalign{\smallskip}
\hfil Source \hfil & $\Theta_{CO}$ \hfil & $\Theta_{{\delta}V}$ \hfil
& $\Theta_{pol}$ \hfil &
$\Theta_{jet} + 90^\circ$ \hfil \cr
\noalign{\smallskip}
\noalign{\medskip}
\noalign{\hrule}
\noalign{\bigskip}
RY Tauri & 48$^\circ\pm5^\circ$ & 21$^\circ$ & ${20^\circ\pm20^\circ}_a$
&  \hfil --- \hfil \cr
DL Tauri & $84^\circ\pm5^\circ$ & $84^\circ$
& ${85^\circ\pm7^\circ}_b$  & \hfil  --- \hfil \cr
DO Tauri & 160$^\circ\pm5^\circ$ & \hfil --- \hfil
& ${170^\circ\pm1^\circ}_a$  & ${160^\circ}_c$ \cr
AS 209 & 138$^\circ\pm2^\circ$ & 120$^\circ$ & ${134^\circ\pm10^\circ}_a$
& \hfil --- \hfil \cr
}
\bigskip
\hrule
\smallskip
\hrule
\bigskip
{\qf a) Bastien (1982)}
\medskip
{\qf b) Tamura \& Sato (1989)}
\medskip
{\qf c) Hirth {\it et al.\ }(1994)}
\bigskip
\vfil
\endinsert
\vfill
\eject

\pageinsert
\vfill
\bigskip
\bigskip
\bigskip
\centerline {\bf Table IV. Derived Keplerian Disk Parameters}
\bigskip
\hrule
\smallskip
\hrule
\medskip
\tabskip=1em plus2em minus0.5em
\halign to \hsize
{\noindent
      # & \hfil # & \hfil # & \hfil # \cr
\noalign{\medskip}
\hfil (1) \hfil & (2) \hfil & (3) \hfil & (4) \hfil \cr
\noalign{\smallskip}
\hfil Source \hfil & $\Delta$V$_{spectra}$ \hfil &  $\Delta$V$_{Kepler}$
\hfil & $i$ \hfil \cr
\noalign{\smallskip}
        &  (km/s) \hfil &(km/s) \hfil & (degrees) \hfil \cr
\noalign{\medskip}
\noalign{\hrule}
\noalign{\bigskip}
RY Tauri & 1.5 & 3.7 & 25  \cr
DL Tauri & 0.3 & 1.4  & 12  \cr
DO Tauri & 1.0 & 1.4  & 45 \cr
AS 209 & 1.5 & 1.8  & 56 \cr
}
\bigskip
\hrule
\smallskip
\hrule



\vfill
\endinsert

\vfill
\eject

\singlespace

\bigskip
\figure{1}
{ a) A map of \trans2\ emission toward RY Tau
integrated over the velocity range, $V_{lsr}$ = 4.2 to 11.1 km s$^{-1}$.
The contour interval is 400 mJy beam$^{-1}$ km s$^{-1}$ (1$\sigma$).
Contours start at the 2$\sigma$ level.
Peak emission is 3.9 Jy beam$^{-1}$ km s$^{-1}$
with an integrated intensity of 7.3 Jy beam$^{-1}$ km s$^{-1}$ .
The synthesized beam, $2\arcsec.6 \times 1\arcsec.9$ FWHM at PA -54.27$^\circ$
with area $\Omega_B = 3.6 \times 10^{-10}$ sr,
is displayed as a hatched region. The apparent elongation to the SE is
largely a beam effect. The FWHM deconvolved size of a Gaussian
fit to the emission is $2\arcsec.3 \times 1\arcsec.9$ oriented
at PA 48.6$^\circ$,
but this value is highly uncertain as an indicator of source size and
orientation, since it is of order the beam size.
b) A map of the mean velocity variations across the structure
seen in a). Grey-scale values correspond to the
velocity scale shown to the right. The contour interval is 0.5 \kms..
A systematic velocity gradient is evident along the axis of elongation.
At the position of peak emission, the gradient is directed along
PA $\sim 21^\circ$.}

\bigskip
\figure{2}
{  a) A map of CO($\2to1$) emission toward DL Tau
integrated over the velocity range, $V_{lsr}$ = 6.5 to 7.7 km s$^{-1}$.
The contour interval is 73 mJy beam$^{-1}$ km s$^{-1}$ (1$\sigma$).
Contours start at the 2$\sigma$ level.
Peak emission is 594 mJy beam$^{-1}$ km s$^{-1}$ with an integrated
intensity of 1.1 Jy beam$^{-1}$ \kms..
The synthesized beam, $2\arcsec.9 \times 2\arcsec.8$
at PA $-85^\circ$ with area $\Omega_B = 6.0 \times 10^{-10}$ sr,
is displayed as a hatched region.
The FWHM deconvolved size of a Gaussian
fit to the emission is $3\arcsec.5 \times 2\arcsec.5$ oriented
at PA 84$^\circ$,
b) A map of the mean velocity variations across the structure
seen in Fig.\ 2a.
Grey-scale values correspond to the
velocity scale at the top. The contour interval is 0.1 \kms..
A gradient is evident along the axis of elongation, at
PA \about 85$^\circ$.}

\bigskip
\figure{3}
{ a) A map of  CO($\2to1$) emission toward DO Tau
integrated over the velocity range, $V_{lsr}$ = -4.8. to 8.2 km s$^{-1}$.
The contour interval is 2.0 Jy beam$^{-1}$ km s$^{-1}$ (2$\sigma$).
Contours start at the 2$\sigma$ level.
Peak emission is 34.5 Jy beam$^{-1}$ km s$^{-1}$
with an integrated intensity of 100.2 Jy beam$^{-1}$ km s$^{-1}$ .
The synthesized beam, $3\arcsec.2 \times 2\arcsec.9$ FWHM at PA -54.27$^\circ$
with area $\Omega_B = 6.9 \times 10^{-10}$ sr,
is displayed as a hatched region. The FWHM deconvolved size of a Gaussian
fit to the emission is $5\arcsec.1 \times 3\arcsec.9$ oriented
at PA 160$^\circ$. The aspect ratio of the major and minor axes
implies an inclination angle of 40$^\circ$ for a circularly symmetric
disk.
b) A map of the mean velocity variations across the structure
seen in a). Grey-scale values correspond to the
velocity scale shown to the right. The contour interval is 0.5 \kms..
A systematic velocity gradient is
not easily identified. This may reflect a combination of kinematic
effects, such as infall, rotation, and outflow.}

\bigskip
\figure{4}
{ a) A map of CO($\2to1$) emission toward AS 209
integrated over the velocity range, $V_{hel}$ = -11.1 to -2.1 km s$^{-1}$.
The contour interval is 350 mJy beam$^{-1}$ km s$^{-1}$, 1$\sigma$.
Contours start at the 2$\sigma$ level.
Peak emission is 3.9 Jy beam$^{-1}$ km s$^{-1}$
with an integrated intensity of 7.4 Jy beam$^{-1}$ km s$^{-1}$ .
The synthesized beam, $3\arcsec.6 \times 2\arcsec.3$ FWHM at PA -18.6$^\circ$
with area $\Omega_B = 6.1 \times 10^{-10}$ sr,
is displayed as a hatched region. The FWHM deconvolved size of a Gaussian
fit to the emission is $3\arcsec.1 \times 2\arcsec.2$ oriented
at PA 137.9$^\circ$. Low level contours are also elongated nearly
perpendicular to this axis, as might be expected for CO entrained
in an outflow.
b) A map of the mean velocity variations across the structure
seen in a). Grey-scale values correspond to the
velocity scale shown to the right. The contour interval is 0.25 \kms..
There is some indication of a velocity gradient along the CO elongation axis.
Deviations from this trend may be due to outflow or infall components.}

\bigskip
\figure{5}
{CO($\2to1$) spectra from the T~Tauri stars, RY Tau,
DL Tau, DO Tau, and AS 209. The integrated intensity from circumstellar
emission in high-spatial-resolution spectral line maps is
plotted as a function of V$_{LSR}$ (V$_{HEL}$ for AS 209).
Variations in velocity resolution are the result of limited bandwidth
coverage for the high velocity resolution channels.
Molecular lineshapes are uniformly plotted over a width of 15 \kms.
for easy comparison.  The continuum flux level,
evident as a perfectly flat line for DL Tau, is
shown outside the region of spectral coverage. The flux scale, in Jy,
is not uniform between spectra, but
ranges over an order of magnitude to facilitate comparison of lineshapes.}

\pageinsert \hugeskip
\vfill
\bigskip
\bigskip
\bigskip
\centerline{\fourteenpoint \bf Figure 1\ \ \ \ \ \ }
\vfill
\endinsert

\pageinsert \hugeskip
\vfill
\bigskip
\bigskip
\bigskip
\centerline{\fourteenpoint \bf Figure 2\ \ \ \ \ \ }
\vfill
\endinsert

\pageinsert \hugeskip
\vfill
\bigskip
\bigskip
\bigskip
\centerline{\fourteenpoint \bf Figure 3\ \ \ \ \ \ }
\vfill
\endinsert

\pageinsert \hugeskip
\vfill
\bigskip
\bigskip
\bigskip
\centerline{\fourteenpoint \bf Figure 4\ \ \ \ \ \ }
\vfill
\endinsert

\pageinsert \hugeskip
\vfill
\bigskip
\bigskip
\bigskip
\centerline{\fourteenpoint \bf Figure 5\ \ \ \ }
\vfill
\endinsert

\end
\pageinsert \hugeskip
\vfil
\bigskip
\bigskip
\bigskip
\bigskip
\caption {\PVDRYT} {Position-Velocity Diagram for RY Tau}
{Position-Velocity Diagrams (PVDs) for RY Tau
taken along axes
parallel (upper right) and perpendicular (lower left) to PA $20^\circ$.
A map of the integrated emission (upper left) has been rotated clockwise
$20^\circ$ for comparison to the PVDs.
This orientation aligns the vertical axis with
the velocity gradient, $\Theta_{\delta{V}}$, seen in Figure 3.1b
and the average linear polarization angle of optical emission
($\Theta_{pol}$). The
deconvolved semi-major axis PA, $\Theta_{CO} \sim 48^\circ$,
is similar but not
well determined because the emission is only marginally resolved.
The apparent elongation direction is largely the result of the shape
of the synthesized beam (see Fig. 3.1a). The diagonal trend of emission
peaks in both PVDs is either the result of combined kinematic effects or
projection of a purely rotational trend on axes which are offset from
the true orientation. The orientation of RY Tau is the most uncertain in
the sample.}
\vfill
\endinsert

\pageinsert \hugeskip
\vfill
\bigskip
\bigskip
\bigskip
\bigskip
\caption {\PVDDLT} {Position-Velocity Diagram for DL Tau}
{Position-Velocity Diagrams (PVDs) for DL Tau taken along axes
parallel (lower left) and perpendicular (upper right) to PA $85^\circ$.
A map of the integrated emission (upper left) has been rotated
counter-clockwise
$5^\circ$ for comparison to the PVDs.
Similar to Fig.\ 3.6, this orientation aligns the horizontal axis with
$\Theta_{CO}$,  $\Theta_{\delta{V}}$ and $\Theta_{pol}$.
A diagonal trend indicating rotation is apparent in the lower left panel.
No similar systematic trend is apparent across the semi-minor axis
(upper right).}
\vfill
\endinsert

\pageinsert \hugeskip
\vfill
\bigskip
\bigskip
\bigskip
\bigskip
\caption {\PVDDOT} {Position-Velocity Diagram for DO Tau}
{Position-Velocity Diagrams (PVDs) for DO Tau
taken along axes
parallel (upper right) and perpendicular (lower left) to PA $160^\circ$.
As in Fig.\ 3.7, a map of the integrated emission (upper left)
has been rotated counter-clockwise
by $20^\circ$ to align the vertical axis with
$\Theta_{CO}$, $\Theta_{pol}$, and an axis perpendicular to $\Theta_{jet}$,
the PA of an optical jet.
At $V_{LSR}$ = 6 \kms., an intensity
minimum separates the blue-shifted emission from a local maximum centered at
the systemic velocity.  This maximum corresponds to the
feature seen in Fig.\ 3.5. The detailed lineshape is not evident
at the velocity resolution used in the PVDs, but
a velocity offset across the minimum
at $V_{LSR}$ = 6 \kms. is most prominent
in the PVD constructed along  the minor axis, confirming
an infall or outflow interpretation
for the blue-shifted emission.}
\vfill
\endinsert

\pageinsert \hugeskip
\vfil
\bigskip
\bigskip
\bigskip
\bigskip
\caption {\PVDAS209} {Position-Velocity Diagram for AS 209}
{Position-Velo-city Diagrams (PVDs) for AS 209
taken along axes
 (upper right) and perpendicular (lower left) to PA $136^\circ$,
in agreement with $\Theta_{CO}$ and $\Theta_{pol}$ from Table 3.3.
As in Fig.\ 3.7, a map of the integrated emission (upper left)
has been rotated clockwise
by $44^\circ$ to align the horizontal axis with
$\Theta_{CO}$ and $\Theta_{pol}$. A diagonal trend indicative of
rotation is evident in the PVD along the semi-major axis (lower left).
Along the minor axis, most emission is centered at the peak position,
but a few displaced peaks are associated with a protrusion from the disk.
The latter may be associated with molecular outflow, but are not oriented
strictly perpendicular to the disk.}
\vfill
\endinsert

\end